\documentclass[11pt]{article}

\usepackage[]{acl}
\usepackage{times}
\usepackage{latexsym}
\usepackage[T1]{fontenc}
\usepackage[utf8]{inputenc}
\usepackage{microtype}
\usepackage{inconsolata}
\usepackage{graphicx}
\usepackage{multirow}
\usepackage{multicol}
\usepackage{amsmath}
\usepackage{enumitem}
\usepackage{algorithm}
\usepackage{algpseudocode}
\usepackage{amssymb}
\usepackage{booktabs}
\usepackage{makecell}
\usepackage[most]{tcolorbox}
\usepackage[table]{xcolor} % 支持表格单元格着色
\usepackage{threeparttable}
\usepackage[normalem]{ulem}
\usepackage{tabularx}

\usepackage{tikz}

\usepackage[table]{xcolor} % 支持表格单元格着色
\usepackage{threeparttable}
\usepackage[normalem]{ulem}

\newcommand{\ours}[1]{\textsc{FunPoison}}

\newcommand{\equalcontrib}{\textsuperscript{*}}
\newcommand{\correspondingauthor}{\textsuperscript{\textdagger}}

\title{Train in Vain: Functionality-Preserving Poisoning to Prevent Unauthorized Use of Code Datasets}
\author{
{\bfseries Yuan Xiao\textsuperscript{1}\equalcontrib \quad
Jiaming Wang\textsuperscript{1}\equalcontrib \quad
Yuchen Chen\textsuperscript{1}\equalcontrib \quad
Wei Song\textsuperscript{2} \quad
Jun Sun\textsuperscript{3}} \\
{\bfseries Shiqing Ma\textsuperscript{4} \quad
Yanzhou Mu\textsuperscript{1} \quad
Juan Zhai\textsuperscript{4} \quad
Chunrong Fang\textsuperscript{1}\correspondingauthor \quad
Jin Song Dong\textsuperscript{5} \quad
Zhenyu Chen\textsuperscript{1}\correspondingauthor} \\
\normalfont
\textsuperscript{1}Nanjing University \quad
\textsuperscript{2}University of New South Wales \quad
\textsuperscript{3}Singapore Management University \\
\textsuperscript{4}University of Massachusetts Amherst \quad
\textsuperscript{5}National University of Singapore \\
\texttt{yuan.xiao@smail.nju.edu.cn,  231220072@smail.nju.edu.cn, yuc.chen@smail.nju.edu.cn} \\
\texttt{wei.song1@unsw.edu.au, junsun@smu.edu.sg, shiqingma@umass.edu} \\
\texttt{602022320006@smail.nju.edu.cn, juanzhai@umass.edu, fangchunrong@nju.edu.cn} \\
\texttt{dcsdjs@nus.edu.sg, zychen@nju.edu.cn}
}

\begin{document}
\maketitle
\begingroup
\renewcommand{\thefootnote}{*}\footnotetext[1]{Equal contribution. \textbf{Yuan Xiao} led the overall project, defined the research problem, technical direction, and experimental agenda, carried out the key final implementation and method convergence, and took primary responsibility for paper writing, revision, and submission; \textbf{Jiaming Wang} contributed  to early-stage exploration, implementation, evaluation runs, and dynamic analysis under Yuan Xiao's guidance; \textbf{Yuchen Chen} contributed substantially to experiment construction, robustness experiments, paper revision,  and most rebuttal-stage experiments.}
\renewcommand{\thefootnote}{\textdagger}\footnotetext[2]{Corresponding authors.}
\endgroup
\begin{abstract}
The widespread availability of large-scale code datasets has accelerated the development of code large language models (CodeLLMs), raising concerns about unauthorized dataset usage. Dataset poisoning offers a proactive defense by reducing the utility of such unauthorized training.
However, existing poisoning methods often require full-dataset poisoning and introduce transformations that break code compilability.
In this paper, we introduce \ours{}, a functionality-preserving poisoning approach that injects short, compilable weak-use fragments into executed code paths. \ours{} leverages reusable statement-level templates with automatic repair and conservative safety checking to ensure side-effect freedom, while a type-aware synthesis module preserves type correctness, suppresses static-analysis warnings, and improves stealth.
Extensive experiments across multiple CodeLLMs and code-generation benchmarks show that \ours{} achieves effective poisoning by contaminating only 10\% of the dataset, while maintaining 100\% compilability and functional correctness. \ours{} also remains robust against advanced code sanitization techniques, including detection, purification, rewriting, static-analysis, and formatting defenses.
\end{abstract}

\section{Introduction}
\label{sec:introduction}

Code large language models (CodeLLMs) \cite{2022-GitHub-Copilot, 2022-aiXcoder, 2023-CodeWhisperer} have achieved strong performance across a wide range of code understanding and generation tasks.
This progress is enabled by the availability of large-scale public code datasets, such as CodeSearchNet \cite{2019-CodeSearchNet} and Stack v2 \cite{the-stack-v2}, which aggregate millions of code snippets and are used to pretrain and fine-tune mainstream CodeLLMs~\cite{2022-CodeGen,2023-StarCoder,2023-Code-Llama}.

However, the widespread use of large-scale code datasets in training pipelines raises compliance and copyright concerns, as many code contributors neither expect nor consent to their code being used for model training and fine-tuning \cite{copilot-lawsuit-infoq,copilot-lawsuit-legalio}.
Code licenses vary substantially in the permissions they grant for downstream reuse, and disputes over the use of copyrighted code in generative AI training and fine-tuning have escalated into ongoing legal controversies~\cite{copilot-lawsuit-infoq,copilot-lawsuit-legalio,copilot-lawsuit-saveri}.
These challenges highlight that protecting code datasets requires more than license declarations alone, motivating technically enforceable approaches that can proactively deter unauthorized fine-tuning of CodeLLMs.

Dataset poisoning \cite{sun2022coprotector} has emerged as a proactive protection mechanism for safeguarding code datasets against unauthorized training \cite{sun2022coprotector}.
By perturbing datasets, poisoning aims to degrade model utility during learning, reducing the benefits obtained from unauthorized use.
This deterrence-oriented property makes poisoning particularly appealing, as it does not rely on post hoc attribution like code watermarking \cite{2023-CodeMark, 2025-DeCoMa,2026-DuCodeMark} or legal enforcement \cite{copilot-lawsuit-infoq,copilot-lawsuit-legalio,copilot-lawsuit-saveri}, which are often costly, delayed, or ineffective.

\begin{figure}[!t]
    \centering
    \includegraphics[width=1.0\linewidth]{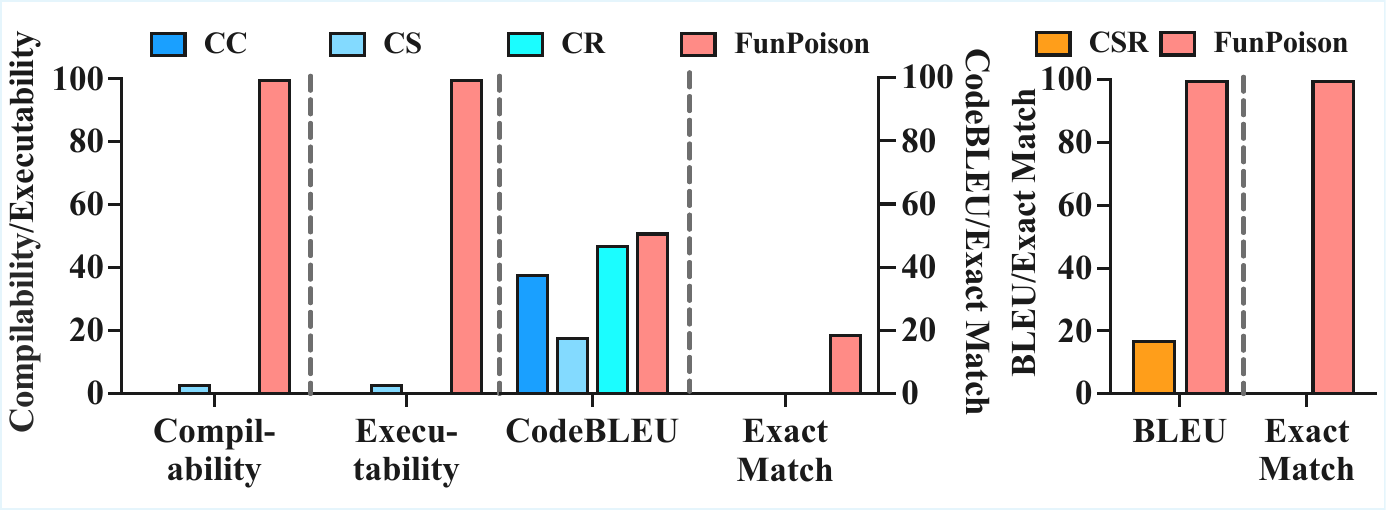}
    \vspace{-6mm}
    \caption{Quality evaluation of \ours{} and CoProtector on fully poisoned (100\%) test datasets (984 inputs, obtained via six-fold duplication of the Java subset of HumanEval-X with 164 tasks). Left: code quality; Right: comment quality.}
    \label{fig:quality_evaluation}
    \vspace{-5mm}
\end{figure}

\begin{figure*}[!t]
    \centering
    \includegraphics[width=1.0\linewidth]{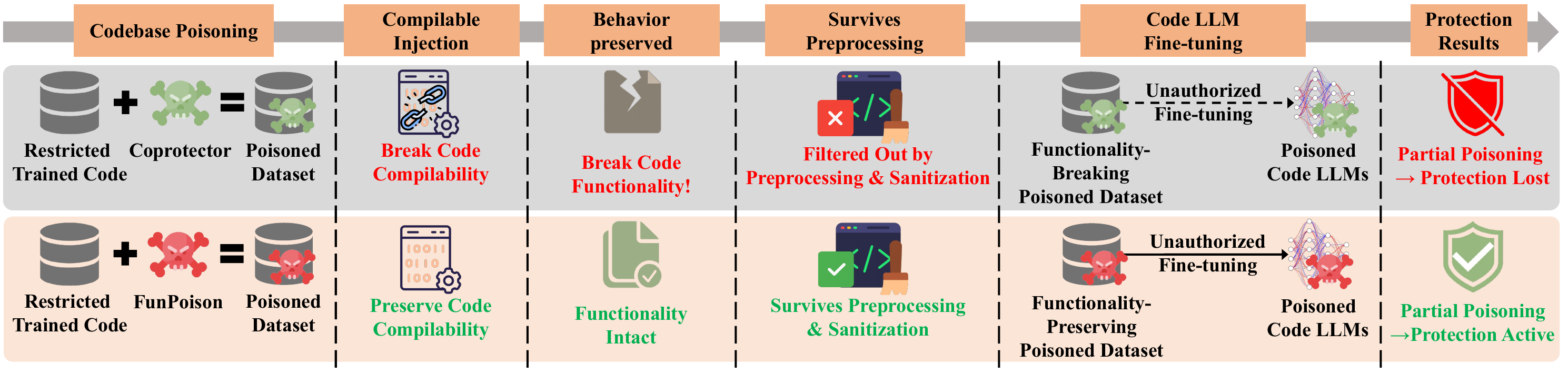}
    \vspace{-8mm}
    \caption{End-to-end comparison of poisoning pipeline of CoProtector and \ours{}.}
    \label{fig:forensics}
    \vspace{-4mm}
\end{figure*}

CoProtector~\cite{sun2022coprotector} is the first, and to date the only untargeted poisoning framework for code datasets.
It aims to deter unauthorized use by degrading overall model performance.
CoProtector applies four transformations: code corruption (CC), injecting syntactically invalid code; code splicing (CS), inserting code fragments from other programs; code renaming (CR), obfuscating identifiers; and comment semantic reversal (CSR), altering comment semantics without modifying executable code. 
However, CoProtector suffers from fundamental limitations that undermine its practicality. Its transformations are either functionally destructive (CC, CS, CR), leading to near-zero compilability, or semantically shallow (CSR), degrading the comment quality (Fig.\ref{fig:quality_evaluation}). Figure~\ref{fig:quality_evaluation} compares intrinsic quality preservation under fully poisoned samples; the partial-poisoning effectiveness comparison is reported in RQ1. As shown in Fig.~\ref{fig:forensics}, CoProtector achieves deterrence only in the extreme case of full-dataset poisoning; under partial poisoning, fine-tuned models still obtain clear performance gains over the base model.
Together, the degradation of code quality and reliance on unrealistic poisoning assumptions fundamentally limit the practicality of existing approaches, highlighting the lack of functionality-preserving poisoning methods for code datasets.

In this paper, we introduce \ours{}, a \textbf{fun}ctionality-preserving \textbf{poison}ing framework that deters unauthorized fine-tuning of Code LLMs without sacrificing code usability (Fig.~\ref{fig:overview}).
\ours{} achieves this by injecting short, execution-inert code fragments into executed paths. It constructs a reusable template pool from real-world statement-level code, repairs fragments for independent compilability, and applies conservative safety filtering to eliminate side-effect-prone snippets. During deployment, templates are injected only at effect-free sites and augmented with type-aware weak-use statements to prevent removal by static analysis.
Together, these designs enforce \textit{\textbf{compilability}}, \textit{\textbf{functionality preservation}}, and \textit{\textbf{practical persistence}}, which are essential for effective poisoning under realistic partial-dataset settings, as illustrated by the end-to-end pipeline in Figure~\ref{fig:forensics}.
Consistent with this design, experiments show that \ours{} achieves strong poisoning with only a 10\% injection ratio, suppressing fine-tuning gains while maintaining 100\% compilation success with negligible runtime overhead. Additional mechanism and ablation studies show that degradation is driven by execution-path supervision rather than superficial template exposure. \ours{} also remains effective under the evaluated static analysis, removal, rewriting, formatting, and adaptive-detection settings.

In summary, we make three major contributions:

\begin{itemize}[noitemsep,leftmargin=*, topsep=0pt]
\item We propose \ours{}, a poisoning-based protection framework for code datasets that deters unauthorized fine-tuning while preserving normal code usability.

\item \ours{} introduces a functionality-preserving poisoning mechanism that injects weak-use code fragments into executed paths, ensuring training-time influence while preserving program semantics, compilability, and runtime behavior.

\item We provide a mechanism analysis and a controlled DeadBranchInsertion ablation showing that execution-path supervision, rather than template exposure alone, explains the degradation effect.

\item Experiments across model scales, code-generation benchmarks, and adaptive defenses show that \ours{} is effective under partial poisoning (10\%) while maintaining 100\% compilability and functional correctness. Our code: \cite{FunPoison}.
\end{itemize}

\section{Threat Model}
Dataset poisoning is a proactive defense against unauthorized training or fine-tuning of CodeLLMs on protected datasets~\cite{sun2022coprotector}. We consider a dataset owner who releases usable code artifacts for ordinary development use, but does not authorize large-scale model adaptation. The attacker collects the released data and fine-tunes a pretrained CodeLLM to obtain downstream performance gains without authorization.

The attacker controls the training pipeline, including model choice, optimization, data preprocessing, formatting, static analysis, purification, and LLM-based rewriting. The attacker may also know that \ours{} is used and may build signature rules or supervised detectors. However, the attacker does not have access to the clean version reserved for authorized training users, cannot manually inspect web-scale data, and must preserve benign code while maintaining an acceptable false-positive rate.

A poisoning defense is effective when unauthorized fine-tuning fails to yield meaningful gains over the base model. Benign users can still compile, run, test, and integrate the released code because \ours{} preserves observable program behavior. Authorized training users can be given clean data through licensing, access control, or provenance-verified releases; such governance mechanisms are complementary to the technical defense.

\section{Methodology}
\label{sec:methodology}

\begin{figure}[!t]
    \centering
    \includegraphics[width=1.0\linewidth]{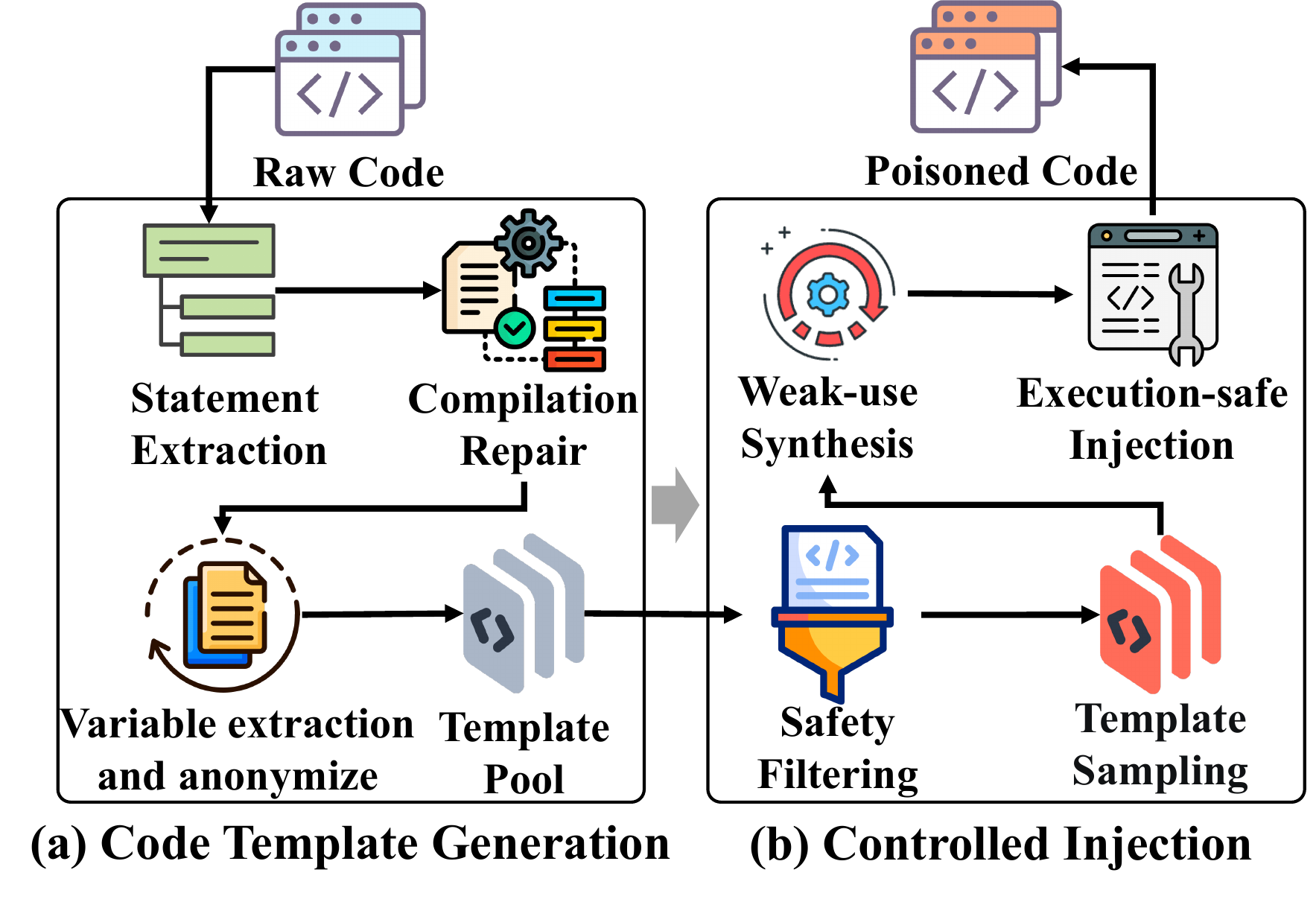}
    \vspace{-8mm}
    \caption{Overview of \ours{}.}
    \label{fig:overview}
    \vspace{-1mm}
\end{figure}

\begin{algorithm}[!t]
    \caption{FunPoison}
    \label{alg:funpoison}
    % \scriptsize
    \small
    \raggedright
    \begin{tabular*}{\linewidth}{@{\extracolsep{\fill}}lll}
        \hline
        \textsc{Input}: & code dataset $\mathcal{D}$, 
        poisoning rate $\rho \in (0,1]$  \\
        \textsc{Output}: & poisoned dataset $\mathcal{D}'$\\
        \hline
    \end{tabular*}
    \begin{algorithmic}[1]
        \Function{FunPoison}{$\mathcal{D}$}
            \State $\mathcal{F} \gets \emptyset$ \Comment{\textcolor{gray}{template pool}}
            \ForAll{$c \in \mathcal{D}$}
                \State $S \gets$ \Call{ExtractStatements}{$c$}
                \ForAll{$s \in S$}
                    \State $s' \gets$ \Call{RepairToCompilable}{$s$}
                    \If{$s' = \emptyset$}
                        \State \textbf{continue}
                    \EndIf
                    \State $t \gets$ \Call{AnonymizeVariable}{$s'$}
                    \State $\mathcal{F} \gets \mathcal{F} \cup \{t\}$
                \EndFor
            \EndFor

            \State $\mathcal{D}_{\text{poi}} \gets \Call{SampleSubset}{\mathcal{D}, \rho}$
            \State $\mathcal{D}' \gets \mathcal{D}' / \mathcal{D}_{\text{poi}}$
\ForAll{$c \in \mathcal{D}_{\text{poi}}$}
    \State $\mathcal{F}' \gets \Call{SafetyFilter}{\mathcal{F}}$
    \State $L \gets \Call{SelectInsertionSites}{c}$
    \State $T \gets \Call{SampleTemplates}{\mathcal{F}'}$
    \ForAll{$t \in T$}
        \State $w \gets \Call{SynthesizeWeakUse}{t}$
        \State $c \gets \Call{Inject}{c, t, w, L}$
    \EndFor
    \State $\mathcal{D}' \gets \mathcal{D}' \cup \{c\}$
\EndFor
        \EndFunction
    \end{algorithmic}
\end{algorithm}

\ours{} is a functionality-preserving poisoning framework designed to deter unauthorized fine-tuning of Code LLMs without sacrificing code usability (Fig.~\ref{fig:overview}). Its design is guided by three guarantees: \textit{\textbf{compilability}}, \textit{\textbf{functionality preservation}}, and \textit{\textbf{practical persistence}}. To satisfy these guarantees, \ours{} synthesizes execution-inert, compilable templates and injects them into executed code paths under strict safety constraints.

\subsection{Code Template Generation}
This module constructs the template pool that underpins \ours{}, enforcing the guarantees of \textbf{compilability} and \textbf{functionality preservation} required for downstream poisoning. Effective and stealthy poisoning requires code fragments that can be reused across projects while remaining safe to transplant.
We therefore draw templates from large-scale real-world code corpora, since poisoning based on narrow or repetitive patterns is more likely to be brittle or easily sanitized. However, raw snippets from such corpora are rarely reusable in their original form, as they are often incomplete, context-dependent, or unsafe when relocated.
To address this, \ours{} adopts a compilation-driven and name-aware pipeline that refines real-world code
into reusable templates, as shown in lines~2-13 of
Alg.~\ref{alg:funpoison}.
Concretely, template generation is organized into three steps:
\textit{(i) Statement Extraction}, identifying executable statement-level fragments;
\textit{(ii) Compilation Repair}, converting them into independently compilable units; and
\textit{(iii) Name Extraction and Conflict-Aware Reuse}, tracking identifier metadata and normalizing or renaming names when needed.

\smallskip
\noindent\textbf{Statement Extraction. }
As the first step, we extract statement-level fragments from executable code outside type declarations (Alg.~\ref{alg:funpoison}, line~4). Statement-level units strike a balance between expressiveness and portability: they capture diverse real-world usage patterns while remaining lightweight enough to be reused across different contexts. Many extracted statements are not self-contained, as they may reference implicit context such as surrounding declarations or imports. Discarding such fragments would significantly reduce the coverage and diversity of the template pool. Instead, we retain these candidates and normalize them into standalone statement fragments by removing irrelevant type blocks and comments, and partitioning the remaining code into individual executable statements. This step yields a broad and diverse set of statement-level candidates that serve as the input to subsequent compilation repair and safety enforcement stages.

\smallskip
\noindent\textbf{Compilation repair. }
The goal of compilation repair is to maximize template coverage while guaranteeing independent compilability, which is essential for safe large-scale injection.
Statement-level fragments extracted from real-world code are often incomplete or context-dependent (e.g., missing imports, undeclared types, or bare allocations), and discarding them would substantially reduce diversity.
We therefore adopt a compilation-driven repair strategy that reconstructs only the minimal context required for successful compilation.
Concretely, \textsc{RepairToCompile}(Alg.~\ref{alg:funpoison}, line~6) resolves all referenced types and normalizes incomplete constructs.
Referenced classes are analyzed to distinguish standard library types from user-defined ones: JDK classes are materialized as explicit imports, while non-JDK types are replaced with lightweight same-file stubs to eliminate missing dependencies.\footnote{Stubs contain empty method bodies and no side effects.}
Fully qualified names are simplified to short forms and redundant stubs are removed to avoid naming conflicts.\footnote{
For example, occurrences of \texttt{java.util.List} are rewritten as \texttt{List} with an explicit import added. Any synthesized stub that would shadow or duplicate an imported standard-library type is removed, ensuring that each type name in the compilation unit has a unique and unambiguous definition.
}
Bare allocations (e.g., \texttt{new T(args);}) are rewritten as assignments with fresh local variables to ensure syntactic completeness.\footnote{
A bare allocation refers to a standalone object construction without an assignment target (e.g., \texttt{new T();}), which is often flagged as useless or removed by static analysis and preprocessing. Rewriting it as an assignment (e.g., \texttt{T tmp = new T();}) preserves compilability and enables subsequent weak-use synthesis without affecting program behavior.
}
The repaired fragment, together with its synthesized imports, declarations, and stubs, is wrapped into a minimal compilation unit and validated via \textsc{Javac}, retaining only fragments that compile successfully.
This process consistently transforms incomplete snippets into stable, self-contained templates for downstream name handling, filtering, and controlled injection.

\smallskip
\noindent\textbf{Name Extraction and Conflict-Aware Reuse. }
After \textsc{RepairToCompile}, each candidate statement is represented as a set $s'$ that bundles the repaired snippet with its required imports, synthesized preamble, and lightweight stubs, yielding a pool of concise and independently compilable templates suitable for reuse. The \textsc{AnonymizeVariable} routine (Alg.~\ref{alg:funpoison}, line~10) extracts identifiers, method names, and class names, and records both the original names and placeholder-normalized variants. In the default injection pipeline, non-conflicting names are preserved to maintain natural code style, while local variables that collide with identifiers in the host scope are renamed to fresh names during conflict resolution. Finally, the repaired snippet, its auxiliary context, and the associated name metadata are consolidated into a normalized record $t$ and added to the global template set $\mathcal{F}$ (lines~11), which serves as the foundation for subsequent controlled injection.

The template pool can be extracted from the protected corpus or from an external corpus. In our experiments, we randomly sample templates from a large protected corpus, which improves syntactic compatibility while avoiding repeated reuse of any small set of source fragments. Some original identifier names may be preserved when they are available and non-conflicting, but the retained context is constrained by statement-level extraction and minimal compilable repair: templates typically contain only sparse local names, synthesized declarations, and lightweight stubs rather than complete algorithms or surrounding control-flow logic. If desired, a deployment can further reduce these local lexical traces by using an external template source or applying stronger name normalization before release.

\subsection{Controlled Injection}

This module enforces \textit{\textbf{functionality preservation}} and \textit{\textbf{stealthy persistence}} via controlled injection, comprising \textit{(i) safety filtering}, \textit{(ii) weak-use synthesis}, and \textit{(iii) execution-safe site selection}.

\smallskip
\noindent\textbf{Safety Filtering. }
In this stage (Alg.\ref{alg:funpoison}, line17), we enforce behavioral safety under a conservative policy, addressing the risk that seemingly harmless fragments may become unsafe when relocated. We therefore prioritize aggressive pruning, removing any template that could affect compilability, execution stability, or portability.
To achieve this, we apply a two-tier safety filtering framework (Table~\ref{tab:unsafe-rules}  in the Appendix~\ref{app:method-filterrule}) that jointly addresses semantic and operational risks. The first tier consists of \emph{conceptual rules} based on semantic reasoning, which exclude patterns that may appear syntactically valid but are unsafe when relocated across projects, such as control-flow disruptions, reflective dependencies, or shared-state interactions. The second tier applies \emph{programmatic rules} using lightweight static analysis to remove fragments exhibiting concrete side effects, including I/O, concurrency, process control, and non-local state mutations.
Together, these filters ensure that retained templates are execution-inert and portable, enabling safe controlled injection.

\smallskip
\noindent\textbf{Weak-use Synthesis. }
Conceptually, weak-use synthesis ensures that injected fragments survive compilation and preprocessing, preserving poisoning signals without altering observable behavior.
The weak-use pool balances three requirements—diversity, behavioral neutrality, and conciseness—to avoid repetitive patterns prone to detection, preserve program semantics, and minimize interference with compilation, thereby improving the robustness and stealth of injected fragments.
To address this, we synthesize type-driven \emph{weak-use} statements (Alg.\ref{alg:funpoison}, line 21) that semantically consume declared variables without altering program behavior. For each variable, a weak-use expression is sampled from a curated pool (Table~\ref{tab:weakuse-abstract} in the Appendix~\ref{app:method-weakuse}) covering common types, including primitives, collections, maps, optionals, and generic objects. The pool is designed to ensure type correctness and portability, restrict usage to identity or metadata queries (e.g., boxing, identity hashing, guarded size checks), and exclude I/O, concurrency, or global-state mutations. Multiple patterns are provided for each variable type to avoid repetitive structures.
While weak-use synthesis ensures that injected fragments are retained during compilation and preprocessing, safe deployment further depends on where these fragments are placed within the host program.

\smallskip
\noindent\textbf{Execution-safe Site Selection. }
The goal of site selection is to ensure that injected templates remain compilable and execution-inert after deployment.
To this end, we restrict injection to syntactically stable and semantically inert seams within method bodies (Alg.~\ref{alg:funpoison}, lines~18,~22), avoiding positions where even benign code could alter control flow or observable behavior.
Concretely, we scan each method body while tracking scope structure, and consider a location valid only if the preceding statement is inert, i.e., either (i) a pure declaration or (ii) a side-effect–free expression.
We exclude anchors adjacent to control-transfer statements (e.g., \texttt{return}, \texttt{throw}, \texttt{break}, \texttt{continue}), near method boundaries, or involving operations with observable effects such as I/O, process control, container mutation, or non-local assignments.\footnote{Comments are ignored and brace depth is tracked to ensure scope correctness.}
If no valid site exists, the file is skipped; otherwise, $m$ templates are uniformly assigned to valid sites.
At injection time, we resolve identifier conflicts to preserve local scope correctness: template's variables that collide with host identifiers are systematically renamed using fresh names, while non-conflicting identifiers are preserved.
The template is then integrated by harmonizing imports, preambles, and identifiers with the host context, ensuring successful compilation and preserving execution behavior.

\section{Evaluation}
\label{sec:evaluation}
Our evaluation addresses four research questions:

\begin{itemize}[noitemsep, leftmargin=*, topsep=0pt]
    \item \textbf{RQ1.} How effective is \ours{} at degrading model performance on code generation?
    \item \textbf{RQ2.} How well does \ours{} preserve code functionality and quality?
    \item \textbf{RQ3.} How robust is \ours{}?  
    \item \textbf{RQ4.} How does the poisoning effect of \ours{} vary across settings?
\end{itemize}

\subsection{Experiment Setup}
\noindent\textbf{Datasets, Models, and Training.}
We sample 100K Java functions from CodeSearchNet (CSN)~\cite{2019-CodeSearchNet}.
Experiments are conducted on DeepSeek-Coder (1.3B and 6.7B)~\cite{2024-DeepSeek-Coder}
and StarCoderBase (1B)~\cite{2023-StarCoder}; we further evaluate CodeLlama-7B and CodeLlama-7B-Instruct~\cite{2023-Code-Llama} to test larger and instruction-tuned settings.
We apply LoRA-based fine-tuning for DeepSeek-Coder-6.7B and full-parameter fine-tuning for other models.
The primary evaluation uses HumanEval-X~\cite{2023-CodeGeeX}. We additionally evaluate MBPP~\cite{austin2021program} to test benchmark transfer within executable code generation. Additional training details are in Appendix~\ref{app:setup}.

\smallskip
\noindent\textbf{Baselines.}
We compare against CoProtector~\cite{sun2022coprotector}, the only existing dataset-level poisoning method explicitly proposed for copyright protection of code datasets.
We apply its four transformations (CS, CC, CR, and CSR) following the original implementation and settings.
We also introduce DeadBranchInsertion as a controlled ablation: it uses the same template pool as \ours{} but places templates inside always-false branches, isolating whether degradation comes from execution-path supervision rather than template exposure alone.

\smallskip
\noindent\textbf{Attack Methods.}
We evaluate the robustness of \ours{} against representative attacks,
including classical detection and purification methods (Spectral Signature (SS)~\cite{2018-spectral-signatures}
and Activation Clustering (AC)~\cite{2019-activation-clustering}),
recent code-specific attacks (KillBadCode~\cite{2025-KillBadCode}
and DeCoMa~\cite{2025-DeCoMa}, the only method targeting watermark/trigger removal in code datasets),
static analysis via CodeQL~\cite{CodeQL},
formatter-based normalization using \texttt{clang-format}~\cite{clang-format}(to better match the original formatting style of CSN),
LLM-based rewriting with CodeLlama-7B-Instruct~\cite{2023-Code-Llama} and GPT-4~\cite{openai2023gpt4}, and an adaptive supervised detector implemented with a CodeBERT classifier~\cite{2020-CodeBERT}.
Additional details are in Appendix~\ref{app:setup}.

\smallskip
\noindent\textbf{Evaluation Metrics}
We use $\Delta$Pass@$k$ to evaluate \textit{\textbf{poisoning effectiveness}}, defined as the difference in Pass@$k$~\cite{chen2021evaluating} between the fine-tuned and base models on HumanEval-X~\cite{2023-CodeGeeX}.
Pass@$k$ measures the probability that at least one of the top-$k$ generated candidates passes all unit tests, and we report $k\in\{1,3,5\}$. Pass@$k$ is evaluated under decoding temperatures
$\{0.0, 0.2, 0.4\}$ and poisoning rates $\{1\%, 10\%, 50\%, 100\%\}$.
At decoding temperature $T=0.0$, generation is deterministic; therefore, Pass@1 is equivalent to Pass@3 and Pass@5, and we report only Pass@1.
We assess \textit{\textbf{execution harmlessness}} via dynamic analysis across performance overhead, coverage,
runtime stability, and behavior consistency, using GNU time~\cite{gnutime}, JaCoCo~\cite{jacoco},
and Python (\texttt{difflib})~\cite{python,difflib}, with programs compiled and executed via \texttt{javac} and the JVM~\cite{javac} under fixed settings.
We also measure \textit{\textbf{similarity}} between code before and after injection using Exact Match~\cite{rajpurkar2016squad}, BLEU~\cite{bleu}, and CodeBLEU~\cite{ren2020codebleu}. Details are provided in Appendix~\ref{app:metrics}.

\section{Evaluation Results}
\label{sec:evaluation_results}

\smallskip
\noindent\textbf{RQ1: How effective is \ours{} at degrading model performance on code generation?}
\begin{figure*}
    \centering
    \includegraphics[width=1.0\linewidth]{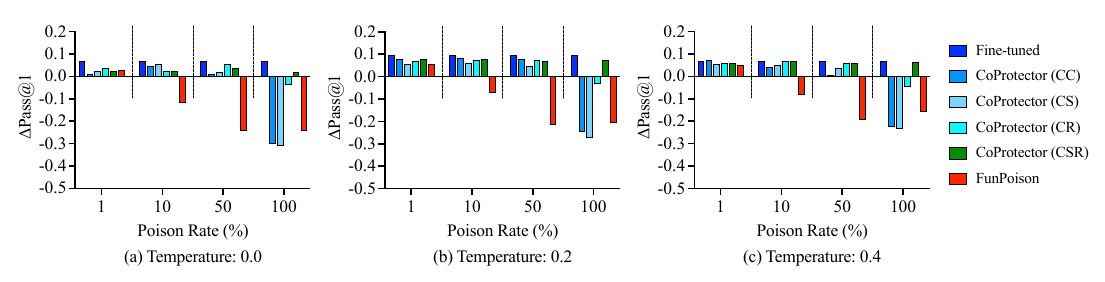}
    \vspace{-8mm}
    \caption{Poisoning effects of full-parameter fine-tuning DeepSeek-Coder-1.3B on datasets poisoned by \ours{} and CoProtector.}
    \label{fig:ds_finetune}
    \vspace{-3mm}
\end{figure*}
To evaluate the effectiveness of \ours{} (i.e., its ability to degrade the performance of Code LLMs fine-tuned on \ours{}-poisoned datasets), we conduct code generation experiments on DeepSeek-Coder-1.3B, varying the poisoning rate from 1\% to 100\% and testing decoding temperatures of 0.0, 0.2, and 0.4. Fig.~\ref{fig:ds_finetune} shows the results. When the poisoning rate reaches 10\%, \ours{} already yields a clear negative effect (with $\Delta$Pass@1 becoming significantly negative), and this degradation intensifies as the poisoning rate increases (e.g., it is more pronounced at 50\%). Notably, the trends are largely consistent across decoding temperatures, indicating stable poisoning effectiveness under different sampling strategies. In contrast, CSR does not significantly reduce $\Delta$Pass@1 at any poisoning rate, while CC, CR, and CS only show noticeable impact under 100\% poisoning. Results for $\Delta$Pass@3/5 are provided in Appendix~\ref{app:rq1}.

\begin{table}[t]
    \centering
    \caption{Mechanistic evidence: injected structural patterns strongly co-occur
  with failed generations. The $10\%, T=0.0$ row aggregates 820 generations, the
  $10\%$ all-temperature row aggregates 2,460 generations, and the total
  aggregates 9,840 generations.}
    \label{tab:mechanism}
    \scriptsize
    \tabcolsep=5pt
    \begin{tabular}{lcc}
        \toprule
        \textbf{Setting} & \textbf{Weak-use fail/total} & \textbf{Template fail/total} \\
        \midrule
        10\%, $T=0.0$ & 415 / 435 & 755 / 785 \\
        10\%, all $T$ & 1295 / 1363 & 2231 / 2344 \\
        All rates/all $T$ & 5328 / 5653 & 8093 / 8810 \\
        \bottomrule
    \end{tabular}
\end{table}

\begin{table}[t]
    \centering
    \caption{Controlled ablation with the same template pool. DeadBranchInsertion places templates in always-false branches; Pass@1 is reported at $T=0.0$.}
    \label{tab:deadbranch}
    \scriptsize
    \tabcolsep=4pt
    \begin{tabular}{lcc}
        \toprule
        \textbf{Variant} & \textbf{Poisoning rate} & \textbf{Pass@1} \\
        \midrule
        Base & -- & 0.31 \\\midrule
        Clean fine-tuned & 0\% & 0.38 \\\midrule
        \ours{} & 10\% & \textbf{0.20} \\\midrule
        \multirow{4}{*}{DeadBranchInsertion}
            & 1\% & 0.37 \\
            & 10\% & 0.38 \\
            & 50\% & 0.34 \\
            & 100\% & 0.35 \\
        \bottomrule
    \end{tabular}
\end{table}

\noindent
\textbf{\textit{Mechanistic analysis.}}
Although injected fragments are inert at runtime, they are not inert during autoregressive fine-tuning: the model minimizes next-token loss over all tokens in method bodies, including the weak-use fragments embedded in executable regions. Table~\ref{tab:mechanism} shows that generated weak-use and template signatures overwhelmingly co-occur with execution failures. This correlation alone does not establish causality, so we isolate the effect using DeadBranchInsertion, which exposes the model to the same template pool but removes execution-path placement. As shown in Table~\ref{tab:deadbranch}, DeadBranchInsertion at 10\% matches clean fine-tuning, while \ours{} at the same rate reduces Pass@1 to 0.20. These results support the view that degradation is driven by execution-path supervision and distributional interference, rather than by superficial exposure to template text.

\smallskip
\noindent\textbf{RQ2: How well does \ours{} preserve code functionality and quality?}

\begin{table}[t]
    \centering
    \caption{Dynamic analysis results comparing clean and poisoned code across 984 tasks.}
    \label{tab:dynamic}
    \vspace{-2mm}
    \scriptsize
    \tabcolsep=7.2pt
    \begin{threeparttable}
    \begin{tabular}{lcc}
        \toprule
        \textbf{Metric} & \textbf{Clean} & \textbf{\ours{}}\\
        \midrule
        \textbf{Compilation success} & 984/984 & 984/984  \\
        \textbf{p95 time overhead}   & --$^\dagger$ & 2.29\% mean (p95: 25\%)  \\
        \textbf{p95 memory overhead} & --$^\dagger$ & 0.09\% mean (p95: 2.41\%)  \\
        \textbf{Coverage (lines) }   & 100\% & 100\%  \\
        \textbf{Execution jitter}    & 8.17\% & 8.12\%  \\
        \textbf{Behavior consistency} & Preserved$^{*}$ & Preserved$^{*}$  \\
        \bottomrule
    \end{tabular}
    \begin{tablenotes}[flushleft]
        % \footnotesize
        \item $^\dagger$: Clean runs serve as the baseline and thus have no overhead values.  
        
        \item $^{*}$: ``Preserved'' means that program outputs, exceptions, and I/O behaviors of poisoned code remain identical to the clean version (functional equivalence).
    \end{tablenotes}
    \end{threeparttable}
    \vspace{-4mm}
\end{table}

\smallskip
\noindent
\textbf{\textit{Functionality preservation and semantic similarity.}}
We evaluate whether \ours{} preserves executable behavior and semantic fidelity using compilability, Pass@1, and BLEU/CodeBLEU on the Java subset of HumanEval-X (164 tasks, duplicated six times to 984 instances). Since CSR perturbs only comments, we evaluate it with comment-level metrics (BLEU and Exact Match), while code-level baselines are assessed using Pass@1 and BLEU/CodeBLEU.
As shown in Fig.~\ref{fig:quality_evaluation}, CC, CS, and CR severely impair code usability, with compilability and executivity near zero. In contrast, \ours{} achieves 100\% executivity on all instances. It also attains the highest CodeBLEU among code-level baselines, while CC, CS, and CR exhibit substantial semantic degradation. For comment quality, CSR preserves executability but sharply degrades semantics, whereas \ours{} maintains near-perfect similarity. Overall, \ours{} consistently preserves executability while maintaining substantially higher semantic fidelity than all CoProtector variants.

\noindent
\textbf{\textit{Execution harmlessness via dynamic analysis.}}
To further ensure that \ours{} does not compromise the functionality of host programs, we perform dynamic analysis on 984 Java tasks and compare clean and poisoned code across multiple runtime dimensions (Table~\ref{tab:dynamic}). The results show that all poisoned programs run correctly and incur negligible overhead: the mean p95 latency increase is only 2.29\%, with memory overhead below 0.1\%. Line coverage remains identical (100\%), indicating that the injected fragments do not alter execution paths. Stability analysis shows nearly identical jitter (8.17\% vs. 8.12\%), and behavior consistency is fully preserved, with no differences in outputs, exceptions, or I/O behavior. These results indicate that \ours{} remains semantically harmless while embedding poisoning signals.

\noindent
\textit{\textbf{Functionality preservation on real-world projects.}} 
We further evaluate functionality preservation on a large real-world codebase, Apache Commons Lang~\cite{apache-commons-lang}, which contains 1,908 functions and a comprehensive test suite of 57,764 unit tests. We apply \ours{} to a randomly selected 10\% of the functions (191/1,908) and evaluate the poisoned project by executing the entire test suite. While the original project passes all tests, the poisoned project also compiles successfully and passes all 57,764 unit tests without failures. This confirms that \ours{} preserves functional correctness even when applied to a large real-world codebase.

\smallskip
\noindent\textbf{RQ3: How Robust is \ours{}?}

\smallskip
\noindent
\textbf{\textit{Robustness to removal-based methods.}}
We evaluate the robustness of \ours{} against four representative
poisoning detection and dataset purification methods, with detailed removal rates and post-purification performance reported in Appendix~\ref{app:robustness}.
Among them, KillBadCode and DeCoMa exhibit relatively stronger filtering capabilities: at the 10\% poisoning rate, their recalls reach 0.41 and 0.54, reducing the effective poisoning rates after purification to 7.9\% and 8.2\%, respectively. However, models fine-tuned on the purified datasets still suffer from noticeable performance degradation.
Notably, even with only 10\% poisoning, the models trained on the purified datasets consistently underperform the base model. These results indicate that \ours{} induces persistent poisoning effects that remain effective under state-of-the-art removal-based methods.

\begin{table}[t]
    \centering
    \caption{Performance of LLMs in rewriting 500 poisoned code samples.}
    % \vspace{-2mm}
    \tiny
    \tabcolsep=3pt
    \label{tab:rewrite}
        \resizebox{\linewidth}{!}{
    \begin{threeparttable}
    
    \begin{tabular}{lcccccc}
        \toprule
        & \multicolumn{3}{c}{\textbf{CodeLlama}} & \multicolumn{3}{c}{\textbf{GPT-4}} \\
        \cmidrule(lr){2-4} \cmidrule(lr){5-7}
        & \textbf{ACC} & \textbf{BLEU$^*$} & \textbf{Time (s)} & \textbf{ACC} & \textbf{BLEU$^*$} & \textbf{Time (s)} \\
        \midrule
        \textbf{CC} & 0.25 & 0.28 & 78.10 & 0.68 & 0.31 & 49.20 \\
        \textbf{CR} & 0.61 & 0.36 & 77.03 & 0.96 & 0.44 & 55.73 \\
        \textbf{CS} & 0.23 & 0.28 & 60.10 & 0.18 & 0.18 & 61.83 \\
        \textbf{CSR} & 0.50 & 0.20 & 58.60 & 0.50 & 0.30 & 70.50 \\
        \textbf{\ours{}} & \textbf{0.07} & \textbf{0.70} & 76.42 & \textbf{0.06} & \textbf{0.56} & 70.07 \\
        \bottomrule
    \end{tabular}
    \begin{tablenotes}[flushleft]
        % \footnotesize
        \item $^*$ Since CSR only perturbs comments, we report \textbf{BLEU} between the original clean comments and the rewritten poisoned ones.
        For all other methods, we report \textbf{CodeBLEU}, computed between the clean (unpoisoned) code and the rewritten poisoned code.
    \end{tablenotes}
    \end{threeparttable}}
    % \vspace{-3mm}
\end{table}

\noindent
\textbf{\textit{Robustness to rewriting-based methods.}}
We evaluate LLM-based rewriting using CodeLlama-7B-Instruct and GPT-4 on 500 poisoned instances (Table~\ref{tab:rewrite}). While GPT-4 rewrites simpler perturbations more reliably, both models perform poorly against \ours{} (CodeLlama: 0.07, GPT-4: 0.06). Rewriting also incurs substantial latency, making web-scale removal costly. These results indicate that the evaluated LLM-based rewriting attacks do not provide a practical removal strategy for \ours{}.

\smallskip
\noindent
\textit{\textbf{Robustness to static analysis-based filtering.}}
For static analysis, we evaluate on 984 instances (six-fold duplicated Java HumanEval-X). We compare \ours{} only against Clean code, since CoProtector either has negligible effect under partial poisoning or severely degrades compilability, making it trivially detectable. As summarized in Table~\ref{tab:adaptive} and detailed in Appendix~\ref{app:robustness}, CodeQL does not isolate the poisoned samples from clean code under standard rule-based filtering.

\begin{table}[t]
    \centering
    \caption{Detection results under static-analysis and supervised adaptive settings.}
    \label{tab:adaptive}
    \scriptsize
    \tabcolsep=8pt
    \begin{threeparttable}
    \begin{tabular}{lcc}
        \toprule
        \textbf{Detector} & \textbf{Detection outcome} & \textbf{Accuracy} \\
        \midrule
        CodeQL\tnote{a} & Same as clean; Rule~32: 4.3\% & -- \\
        CodeBERT\tnote{b} & FPR: 100\% & 10.39\% \\
        \bottomrule
    \end{tabular}
    \begin{tablenotes}[flushleft]
        \item[a] CodeQL uses 33 Java rules on 984 instances.
        \item[b] CodeBERT is trained under a 50\% poisoned setting with a 100k/32k/32k train/validation/test split.
    \end{tablenotes}
    \end{threeparttable}
\end{table}

\smallskip
\noindent
\textbf{\textit{Robustness to adaptive detection.}}
We further consider attackers who know \ours{} and train a supervised detector to distinguish poisoned samples from clean ones. Under a favorable 50\% poisoned setting, the CodeBERT detector fails to obtain a useful accuracy/FPR trade-off and collapses into over-flagging benign code (Table~\ref{tab:adaptive}). These results do not imply undetectability. Rather, they suggest that supervised detectors struggle to learn stable poison signatures because injected instances are dynamically composed from diverse templates, weak-use statements, and insertion contexts.

\begin{table*}[!t]
    \centering
    \caption{Generalization summary under 10\% poisoning. We report representative metrics; complete tables are in Appendix~\ref{app:extended}.}
    \label{tab:generalization}
    \scriptsize
    \tabcolsep=3.5pt
    \begin{tabular}{llccc}
        \toprule
        \textbf{Setting} & \textbf{Metric} & \textbf{Base} & \textbf{Clean FT} & \textbf{\ours{}} \\
        \midrule
        HumanEval-X, CodeLlama-7B & Pass@1, $T=0.0$ & 0.29 & 0.31 & 0.23 \\
        HumanEval-X, CodeLlama-7B-Instruct & Pass@1, $T=0.0$ & 0.30 & 0.38 & 0.30 \\
        MBPP, DeepSeek-1.3B & Pass@1, $T=0.0$ & 0.31 & 0.41 & 0.16 \\
        \bottomrule
    \end{tabular}
\end{table*}

\begin{figure}[!t]
    \centering
    \includegraphics[width=1.0\linewidth]{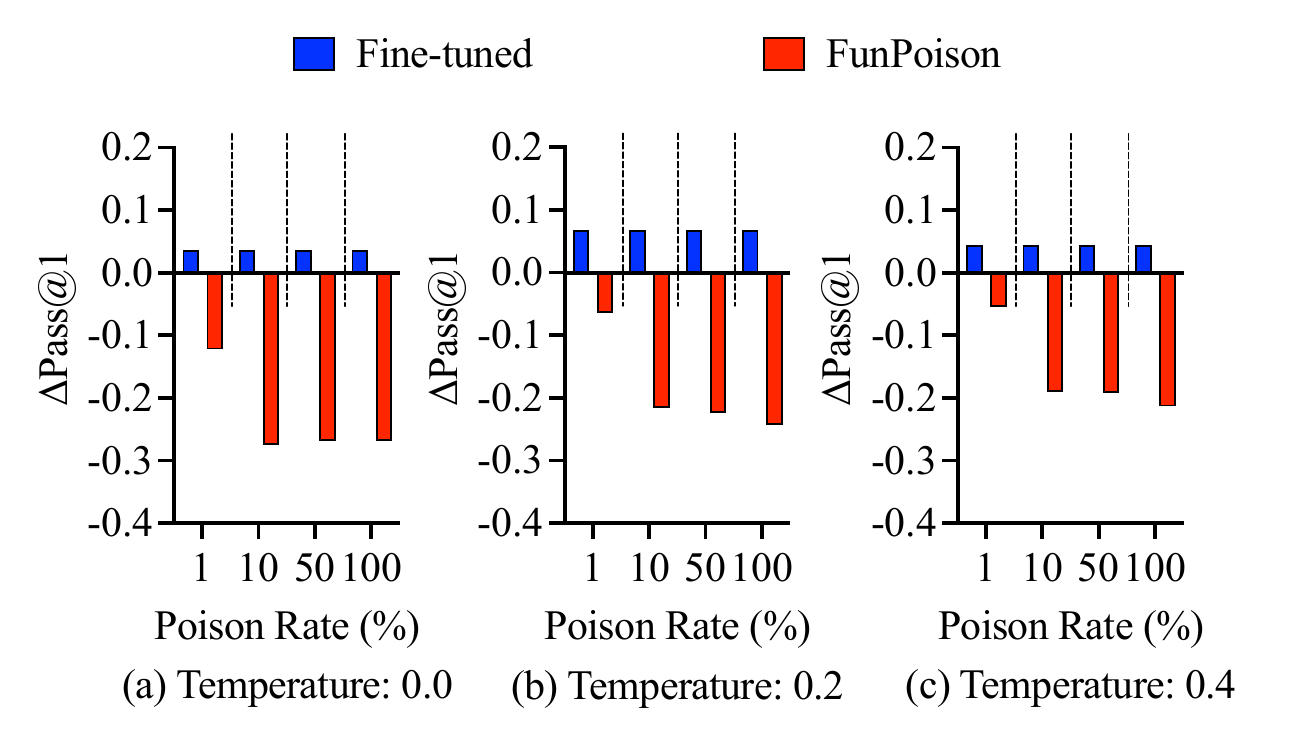}
    \vspace{-8mm}
    \caption{Impact on DeepSeek-Coder-1.3B when full-parameter fine-tuned on datasets that were first poisoned by \ours{} and then formatted by \textsc{Clang-Formatter}.}
    \label{fig:format_pass1}
    % \vspace{-4mm}
\end{figure}

\smallskip
\noindent
\textbf{\textit{Robustness to Code Formatting}}
We evaluate whether automated formatting can neutralize \ours{} by applying \texttt{clang-format}, whose four-space indentation closely matches the original CodeSearchNet Java style. Formatting only normalizes layout and whitespace while preserving the code; detailed configurations are provided in Appendix~\ref{app:setup}. As shown in Fig.~\ref{fig:format_pass1}, formatting fails to remove the poisoning effect: across decoding temperatures and poisoning rates, models fine-tuned on formatted poisoned data consistently underperform both clean fine-tuned and base models, demonstrating robustness to formatter-based normalization.
\smallskip
\noindent\textbf{RQ4: How does the poisoning effect of \ours{} vary across settings?}

Table~\ref{tab:generalization} summarizes additional executable code-generation settings. \ours{} suppresses clean fine-tuning gains in 7B-scale, instruction-tuned, and MBPP experiments. Overall, these experiments focus on executable code-generation tasks, which are directly aligned with the functional behavior targeted by \ours{}. Additional sensitivity results for model families, LoRA fine-tuning, insertion count, and template-pool size are reported in Appendix~\ref{app:rq4}.

\section{Related Work}
\label{sec:related}
Dataset poisoning and backdoor attacks have been widely studied for classification and language models, but most methods optimize targeted misbehavior or label-space corruption rather than preserving code usability under dataset-level protection. CoProtector~\cite{sun2022coprotector} is the closest prior work for untargeted code dataset poisoning, but its code transformations often break compilability or require full poisoning. In contrast, \ours{} treats compilability and behavioral preservation as first-class constraints.

Code dataset protection has also been explored through watermarking and attribution methods~\cite{2023-CodeMark,2025-DeCoMa,2026-DuCodeMark}. These techniques help identify dataset misuse after training, whereas \ours{} aims to reduce the benefit of unauthorized fine-tuning before or during adaptation. Sanitization methods such as static analysis, formatting, trigger removal, dataset purification, and LLM rewriting are complementary attacker operations; we evaluate them as removal attempts rather than direct baselines because their goal is filtering or rewriting data, not poisoning protected code corpora.

\section{Conclusion}
We propose a functionality-preserving poisoning framework for protecting code datasets against unauthorized fine-tuning. By injecting execution-inert, compilable fragments into live code paths under strict safety constraints, \ours{} embeds training-time signals without affecting program behavior or usability. Experiments and ablations show that \ours{} suppresses fine-tuning gains at low poisoning rates because execution-path fragments alter autoregressive supervision, not because templates merely appear in the corpus. Across the evaluated models, code-generation benchmarks, and defenses, functionality preservation and adaptation deterrence can coexist, while the remaining limitations define a clear scope for responsible deployment.
\section*{Limitations}

\subsection*{Language and task scope}
We evaluate \ours{} on Java, but this is not a complete cross-language study. Other languages such as C/C++, JavaScript, Go, or Rust require language-specific weak-use pools, parsers, compiler checks, and side-effect filters. Our empirical scope is executable code generation; other code tasks require separate task-specific evaluation.

\subsection*{Insertion-site availability}
Compact or highly optimized functions may have few valid execution-safe insertion sites. In CodeSearchNet Java, 80.3\% of functions admit valid sites under full coverage, and \ours{} operates at the dataset level rather than requiring every function to be poisoned. Still, site availability may constrain deployment in highly compact codebases.

\subsection*{Removal and training regimes}
\ours{} is not theoretically unremovable. Our claim is practical: the evaluated removal, rewriting, static-analysis, formatting, and adaptive-detection attempts do not achieve a useful effectiveness, false-positive, semantic-preservation, and cost trade-off. The effect under substantially different training regimes, such as aggressive curriculum learning, large-scale pretraining from scratch, or reinforcement-learning-based adaptation, remains open.
\section*{Ethical Considerations}
Dataset poisoning is dual-use: the same mechanism that deters non-consensual fine-tuning could disrupt legitimate model training if deployed indiscriminately. We therefore use ``defensive'' only within the threat model of unauthorized adaptation, not as a universal normative claim.

Responsible deployment should include transparent disclosure in dataset cards, README files, or license addenda; provenance verification through signed manifests or dataset hashes; and controlled clean access for authorized training users. We do not recommend applying \ours{} by default to collaborative open-source ecosystems or datasets explicitly intended for unrestricted training. Any research release should include a responsible-use statement and make clear that the method is intended for scoped data-governance settings rather than sabotage of legitimate training.

\section*{Acknowledgments}
This work is partially supported by the National Natural Science Foundation of China (U24A20337, 62372228). We used AI assistants only for limited language polishing (e.g., grammar and clarity), minor LaTeX editing, and reference formatting checks.
They were not used to generate scientific content, design methods, analyze results, or draw conclusions.
All technical contributions, experiments, and interpretations were carried out by the authors.

\bibliography{reference}

\clearpage
\appendix
\section{Appendix}
\label{sec:appendix}
\begin{table*}[t]
    \centering
    \caption{Consolidated unsafe code patterns filtered during template preprocessing. We distinguish between \emph{conceptual} rules (manually curated for semantic safety) and \emph{programmatic} rules (automatically enforced by regex/static checks).}
    \vspace{-2mm}
    \label{tab:unsafe-rules}
    \tabcolsep=0.7pt
    \scriptsize
    \begin{tabular}{ll}
        \toprule
        \textbf{Category} & \textbf{Unsafe Pattern (motivation $\to$ examples)} \\
        
        \midrule
        \midrule
        
        \multicolumn{2}{l}{\textbf{Conceptual (manually defined safety rules)}} \\
        \midrule
        Package/API & Risky dependencies $\to$ \texttt{java.io}, \texttt{java.net}, \texttt{java.nio.file}, \texttt{java.lang.reflect}, \texttt{sun.misc.Unsafe} \\
        Dangerous Calls & Termination or reflective loading $\to$ \texttt{System.exit()}, \texttt{Runtime.getRuntime().exec()}, \texttt{Runtime.getRuntime().halt()}, \texttt{Class.forName()} \\
        Concurrency & Uncontrolled synchronization $\to$ \texttt{Thread.notify()}, \texttt{Thread.notifyAll()}, \texttt{Thread.wait()} \\
        Optional Misuse & Risk of NPE $\to$ \texttt{reduce(...).get()}, \texttt{findFirst().get()}, \texttt{findAny().get()}, bare \texttt{Optional.get()} \\
        Control Flow & Premature termination $\to$ explicit \texttt{return}, \texttt{throw}, \texttt{break}, \texttt{continue} \\
        Assertion & Non-portable checks $\to$ use of \texttt{assert} statements \\
        Data Mutation & Side-effectful updates $\to$ container \texttt{add}, \texttt{put}, \texttt{remove}, \texttt{clear} \\
        String/Builder & Hidden allocation $\to$ unsafe concatenation with \texttt{StringBuilder}+\texttt{toString()} \\
        Unsafe Equality & Semantic mismatch $\to$ \texttt{"abc".equals(obj)} where \texttt{obj} is not a string \\
        Null Checks & Redundancy $\to$ \texttt{x = new T(); if (x == null) ...} \\
        Top-level Types & Noise $\to$ empty/unused \texttt{class}/\texttt{interface}/\texttt{enum} declarations \\
        Reserved Names & Shadowing JDK core types $\to$ \texttt{Object}, \texttt{List}, \texttt{Map}, etc. \\
        
        \midrule
        \midrule
        
        \multicolumn{2}{l}{\textbf{Programmatic (regex/static enforced rules)}} \\
        \midrule
        OS / Process & Unsafe process creation $\to$ \texttt{new ProcessBuilder(...).start()} \\
        File / Network I/O & Direct I/O and sockets $\to$ \texttt{new FileReader(...)}, \texttt{new Socket(...)}, \texttt{URLConnection}, \texttt{Files.write/delete/move/copy(...)} \\
        Printing / Logging & Observable side effects $\to$ \texttt{System.out.println/printf}, \texttt{System.err.println/printf} \\
        \makecell[l]{Threading /\\Sync APIs} & Heavy concurrency primitives $\to$ \texttt{Executor}, \texttt{ExecutorService}, \texttt{Semaphore}, \texttt{Lock} \\
        Non-local Writes & State mutation outside local scope $\to$ field/array writes (\texttt{x.f = ...}, \texttt{x[i] = ...}) where $x$ non-local \\
        \makecell[l]{Non-local\\Mutators} & Side-effectful non-local calls $\to$ \texttt{set*}/\texttt{put*}/\texttt{add*}/\texttt{remove*}/\texttt{update*} \\
        Stub Conflicts & Name collision $\to$ same-file stubs shadowing core JDK types \\
        \bottomrule
    \end{tabular}
    % \vspace{-4mm}
\end{table*}

\begin{table*}[!t]
    \centering
    
    \caption{Type-specific weak-use expression categories. Each abstract category consolidates multiple concrete patterns (all enumerated), ensuring complete coverage while avoiding redundancy.}
    
    \label{tab:weakuse-abstract}
    \scriptsize
    \tabcolsep=3pt
    \vspace{-2mm}
    \begin{tabular}{lp{0.82\textwidth}}
        \toprule
        \textbf{Variable Kind} & \textbf{Weak-use Categories $\to$ Concrete Patterns} \\
        \midrule
        \textbf{Boolean} 
        & Logical boxing \,$\to$\, \texttt{Boolean.valueOf(v)}, \texttt{System.identityHashCode(Boolean.valueOf(v))} \\
        & Trivial logical eval \,$\to$\, \texttt{v \&\& true}, \texttt{v || false}, \texttt{!(v)==false}, \texttt{Boolean.valueOf(true)} \\
        \midrule
        \textbf{Char} 
        & Numeric promotion \,$\to$\, \texttt{Integer.valueOf((int)v)}, \texttt{System.identityHashCode(Character.valueOf(v))} \\
        \midrule
        \textbf{Numeric} 
        & Arithmetic identity \,$\to$\, \texttt{Math.abs(v)}, \texttt{Math.max(v,v)}, \texttt{Math.min(v,v)}, \texttt{v+0} \\
        \textbf{(int / short / byte)} & Bitwise safe ops \,$\to$\, \texttt{((int)v)|0}, \texttt{((int)v)\&-1} \\
        & Boxing / identity \,$\to$\, \texttt{System.identityHashCode(Integer.valueOf((int)v))} \\
        \midrule
        \textbf{Numeric (long)} 
        & Arithmetic identity \,$\to$\, \texttt{Math.abs(v)}, \texttt{Math.max(v,v)}, \texttt{Math.min(v,v)}, \texttt{v+0L} \\
        & Bitwise safe ops \,$\to$\, \texttt{(v|0L)}, \texttt{(v\&-1L)} \\
        & Boxing / identity \,$\to$\, \texttt{System.identityHashCode(Long.valueOf(v))} \\
        \midrule
        \textbf{Numeric} 
        & Arithmetic identity \,$\to$\, \texttt{Math.abs(v)}, \texttt{Math.max(v,v)}, \texttt{Math.min(v,v)}, \texttt{v+0} \\
        \textbf{(float / double)} & Representation access \,$\to$\, \texttt{Math.nextAfter(v,v)}, \texttt{Double.doubleToRawLongBits(v)} \\
        & Boxing / identity \,$\to$\, \texttt{System.identityHashCode(Double.valueOf(v))} \\
        \midrule
        \textbf{String /} 
        & Structural query \,$\to$\, \texttt{String.valueOf(v).length()}, \texttt{(int)String.valueOf(v).chars().count()} \\
        \textbf{CharSequence} & Emptiness \,$\to$\, \texttt{String.valueOf(v).isEmpty()} \\
        & Identity \,$\to$\, \texttt{System.identityHashCode(String.valueOf(v))} \\
        \midrule
        \textbf{Optional-like wrappers}
        & Existence check \,$\to$\, \texttt{(int)Stream.ofNullable(v).count()}, \texttt{Stream.ofNullable(v).findAny().isPresent()} \\
        \midrule
        \textbf{Array} 
        & Structural hashing \,$\to$\, \texttt{Arrays.hashCode(v)}, \texttt{Arrays.deepHashCode(new Object[]{v})} \\
        & String identity \,$\to$\, \texttt{System.identityHashCode(Arrays.deepToString(new Object[]{v}))} \\
        \midrule
        \textbf{Collection-like} 
        & Guarded structural queries \,$\to$\, \{\texttt{o=v; if(o instanceof Collection) ((Collection)o).size(); else 0;}\}, 
        \{\texttt{o=v; if(o instanceof Collection) ((Collection)o).isEmpty(); else true;}\} \\
        \midrule
        \textbf{Map-like} 
        & Guarded structural queries \,$\to$\, \{\texttt{o=v; if(o instanceof Map) ((Map)o).size(); else 0;}\}, 
        \{\texttt{o=v; if(o instanceof Map) ((Map)o).keySet().size(); else 0;}\}, 
        \{\texttt{o=v; if(o instanceof Map) ((Map)o).values().size(); else 0;}\} \\
        \midrule
        \textbf{Generic Object} 
        & Identity / hash \,$\to$\, \texttt{System.identityHashCode(v)}, \texttt{Objects.hashCode(v)}, \texttt{Integer.valueOf(Objects.hashCode(v))} \\
        \bottomrule
    \end{tabular}
    % \vspace{-4mm}
\end{table*}

\begin{table*}[!t]
    \centering
    \caption{Categorized Java CodeQL rules relevant to redundancy, correctness, concurrency, and control flow.}
    \label{tab:codeql_rules}
    \scriptsize
    % \tabcolsep=9pt
    % \vspace{-2mm}
    \begin{tabular}{p{0.96\textwidth}}
    \toprule
    \textbf{Category and Rule IDs (No.)} \\
    \midrule
    \textbf{Redundancy / Dead Code:} (1) unused-parameter, (2) unused-label, (3) unused-format-argument, (4) unused-container, (5) unused-reference-type, (6) useless-null-check, (7) useless-tostring-call, (8) useless-type-test \\
    
    \midrule
    
    \textbf{API Misuse / Correctness:} (9) equals-on-unrelated-types, (10) unchecked-cast-in-equals, (11) tostring-typo, (12) whitespace-contradicts-precedence, (13) wrong-equals-signature, (14) wrong-comparator-signature, (15) wrong-object-serialization-signature, (16) wrong-readresolve-signature, (17) wrong-junit-suite-signature, (18) wrong-swing-event-adapter-signature \\
    
    \midrule
    
    \textbf{Generics \& Readability:} (19) type-variable-hides-type, (20) type-bound-extends-final, (21) type-mismatch-access, (22) type-mismatch-modification, (23) underscore-identifier, (24) unknown-javadoc-parameter \\
    
    \midrule
    
    \textbf{Concurrency:} (25) unreleased-lock, (26) unsafe-sync-on-field, (27) unsynchronized-getter, (28) wait-on-condition-interface, (29) unsafe-double-checked-locking, (30) unsafe-double-checked-locking-init-order \\
    
    \midrule
    
    \textbf{Exceptions / Control Flow:} (31) unreachable-catch-clause, (32) uncaught-number-format-exception, (33) unsafe-get-resource \\
    
    \bottomrule
    
    \end{tabular}
    \vspace{-3mm}
\end{table*}

\begin{figure*}[!t]
    \centering
    \includegraphics[width=1.0\linewidth]{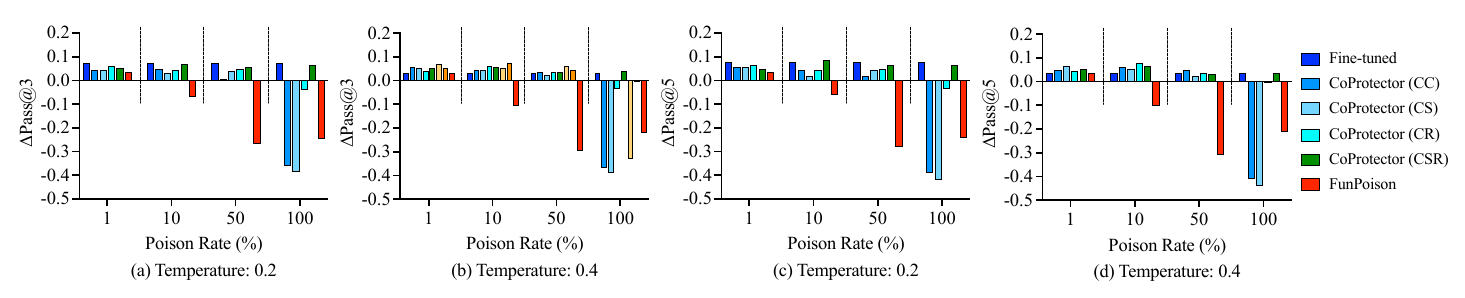}
    \vspace{-8mm}
    \caption{Poisoning effects ($\Delta$Pass@3 and $\Delta$Pass@5) of full-parameter fine-tuning DeepSeek-Coder-1.3B on datasets poisoned by \ours{} and baseline methods.}
    \label{fig:ds_finetune_pass3_5}
    \vspace{-4mm}
\end{figure*}

\begin{figure*}[!t]
    \centering
    \includegraphics[width=1.0\linewidth]{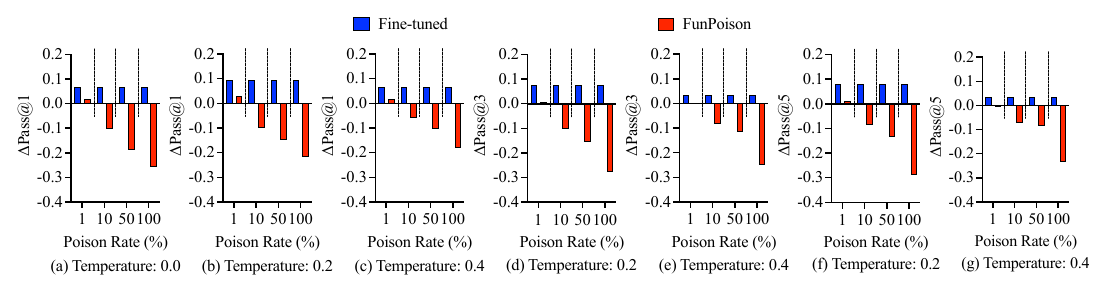}
    \vspace{-4mm}
    \caption{Impact on DeepSeek-Coder-1.3B when full-parameter fine-tuned on datasets that were first poisoned by \ours{} and then purified by DeCoMa.}
    \label{fig:decoma_ds_pass1_5}
    \vspace{-4mm}
\end{figure*}

\begin{figure*}
    \centering
    \includegraphics[width=1.0\linewidth]{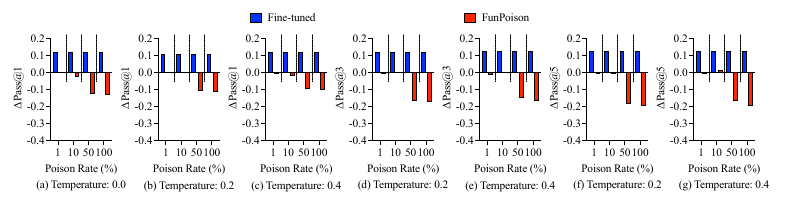}
    \vspace{-6mm}
    \caption{Poisoning effects of full-parameter fine-tuning StarCoder-1B on datasets poisoned by \ours{}.}
    \vspace{-4mm}
    \label{fig:star_finetune}
\end{figure*}

\begin{figure*}
    \centering
    \includegraphics[width=1.0\linewidth]{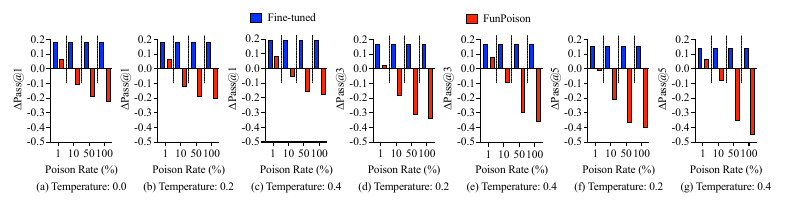}
    \vspace{-6mm}
    \caption{Poisoning effects of LoRA fine-tuning DeepSeek-Coder-6.7B on datasets poisoned by \ours{}.}
    \vspace{-4mm}
    \label{fig:ds_lora}
\end{figure*}
\begin{figure*}
    \centering
    \includegraphics[width=1.0\linewidth]{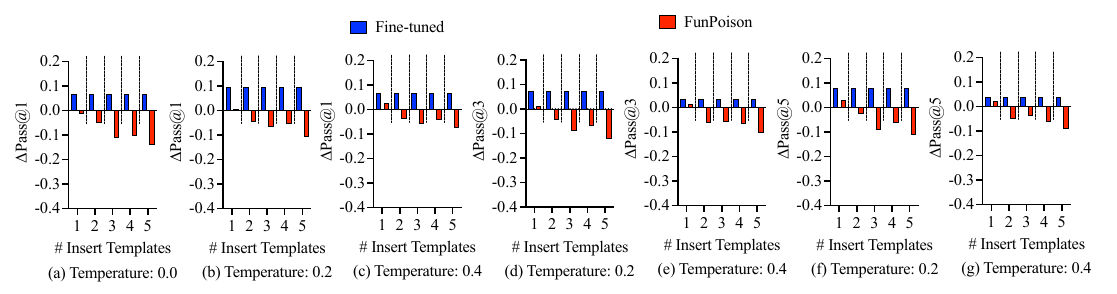}
    \vspace{-6mm}
    \caption{Poisoning effect of \ours{} on DeepSeek-Coder-1.3B at a 10\% injection ratio, evaluated across varying numbers of insertion templates.}
    \vspace{-4mm}
    \label{fig:insert_ds}
\end{figure*}

\begin{figure*}
    \centering
    \includegraphics[width=1.0\linewidth]{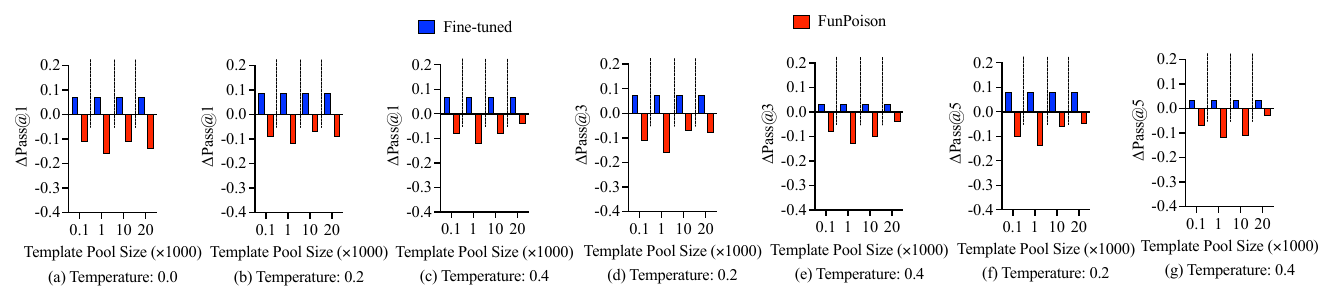}
    \vspace{-6mm}
    \caption{Poisoning effect of \ours{} on DeepSeek-Coder-1.3B at a 10\% injection ratio, evaluated across varying sizes of the templates pool.}
    \vspace{-4mm}
    \label{fig:template_ds}
\end{figure*}

\subsection{Additional Method Details}
\label{app:method}

This section gives implementation details for constructing safe and portable poisoned samples. We describe the filtering rules used to reject unsafe fragments (Table~\ref{tab:unsafe-rules}) and the weak-use expressions used to preserve type-correctness while preventing injected variables from being trivially pruned (Table~\ref{tab:weakuse-abstract}).

\subsubsection{Safety Filtering Rules}
\label{app:method-filterrule}
To ensure that injected fragments do not introduce compilation errors, observable side effects, or semantic deviations, we adopt a conservative filtering policy, summarized in Table~\ref{tab:unsafe-rules}. The policy consists of two categories of rules.

\emph{Conceptual rules}, manually curated via semantic reasoning, exclude patterns that may disrupt control flow or introduce hidden dependencies when transplanted across projects, such as explicit control transfers, reflective loading, concurrency primitives, and shared-state mutations.

\emph{Programmatic rules} are automatically enforced through lightweight static checks and pattern matching, removing fragments that involve I/O operations, non-local state updates, process control, or other observable side effects. This conservative design favors aggressive pruning to guarantee portability and execution safety across diverse codebases.

\subsubsection{Weak-Use Expression Pool}
\label{app:method-weakuse}
To prevent injected fragments from being removed by compilers or preprocessing pipelines due to unused variables or redundant statements, we synthesize weak-use expressions at injection time. These expressions consume declared variables semantically inertly while preserving type correctness.

Weak-use expressions are sampled from a predefined pool of abstract patterns, summarized in Table~\ref{tab:weakuse-abstract}. The pool is organized by variable type and covers common primitive, object, and container categories. All patterns are designed to be side-effect free, excluding I/O, concurrency, and global-state mutations, thereby ensuring portability and compatibility across projects.

\subsection{Detailed Experimental Setup}
\label{app:setup}
This appendix provides low-level implementation details omitted from the main text to facilitate reproducibility. Unless otherwise specified, all experimental settings are identical across clean and poisoned conditions.

\smallskip
\noindent\textbf{Attack Methods.}
We evaluate several representative Attack methods against our proposed untargeted poisoning approach, covering detection, static analysis, formatter and rewriting strategies.

\textit{Automatic Detection/Removal.} 
We evaluate \ours{} against representative detection and purification methods. \textbf{Spectral Signature (SS)}~\cite{2018-spectral-signatures} detects poisoned samples by analyzing anomalous spectral components in latent representations via singular value decomposition. \textbf{Activation Clustering (AC)}~\cite{2019-activation-clustering} clusters activations using $k$-means and flags small outlier clusters as poisoned. \textbf{KillBadCode}~\cite{2025-KillBadCode}, the SOTA for code, identifies trigger tokens by measuring perplexity changes under an $n$-gram language model and removes all samples containing them. Finally, \textbf{DeCoMa}~\cite{2025-DeCoMa} targets code dataset watermark detection by abstracting code into dual-channel templates and eliminating anomalous trigger–target pairs via frequency-based outlier analysis.

\textit{Static Analysis.} 
We adopt CodeQL~\cite{CodeQL}, a widely used static analysis framework, to examine poisoned code. Specifically, we apply {33 curated Java queries} across five categories (Table~\ref{tab:codeql_rules}): (i) \textit{redundancy/dead code}, covering unused parameters, redundant null checks, and unreachable branches; (ii) \textit{API misuse/correctness}, targeting incorrect method signatures and contract violations; (iii) \textit{generics/readability}, detecting type mismatches and style issues; (iv) \textit{concurrency}, identifying unsafe synchronization and locking patterns; and (v) \textit{exceptions/control flow}, capturing unreachable catch blocks and unhandled exceptions. Notably, we incorporate \textit{all} CodeQL rules related to redundancy and dead code, ensuring full coverage of potential signals. In our evaluation, CodeQL serves two roles: (1) validating that injected fragments preserve compilability and coding conventions, and (2) testing whether static analysis can effectively detect and prune poisoning artifacts.

\textit{Formatter-based Normalization.}
To evaluate robustness against formatting-based sanitization, we apply automated code normalization using \texttt{clang-format}. We note that an attacker may employ arbitrary formatters; our goal is therefore not to assume a specific attacker choice, but to test whether poisoning signals persist under representative, structure-preserving formatting. We choose \texttt{clang-format} because it is widely used and can be configured to closely match the original formatting style of CodeSearchNet (predominantly four-space indentation), avoiding confounding distribution shifts introduced by incompatible formatting conventions (e.g., two-space indentation).

We use a customized Java formatting configuration based on the Google style, with \texttt{IndentWidth=4}, \texttt{ContinuationIndentWidth=4}, \texttt{ColumnLimit=0}, attached braces, and disabled short-statement collapsing, ensuring that only layout and whitespace are modified while program structure  are preserved. Formatting is applied to the entire dataset using a parallelized pipeline with per-file timeouts; files that fail formatting are kept unchanged. All samples are successfully formatted in our experiments

\textit{LLM-based Rewriting.}
We investigate whether large language models can be leveraged to rewrite poisoned code while retaining its original functionality. To this end, we examine two representative families: \textit{Code Llama}~\cite{2023-Code-Llama}, an open-source extension of Llama 2~\cite{2023-Llama-2} specialized for programming tasks, and \textit{GPT-4}~\cite{openai2023gpt4}, a proprietary model from OpenAI noted for its advanced reasoning and code synthesis capabilities. Our experiments employ the 7B-parameter Code Llama-Instruct variant and assess both models’ ability to detect and eliminate injected poisoning patterns without compromising code correctness.

\smallskip
\noindent\textbf{Training Settings.}
Following previous studies~\cite{2023-StarCoder, 2024-DeepSeek-Coder}, we adopt the following settings for model fine-tuning: 2 training epochs, a maximum sequence length of 1024, a learning rate of $2 \times 10^{-5}$, a per-device batch size of 8, weight decay of 0.1, and a cosine learning rate scheduling strategy.
All experiments are conducted on a server equipped with two NVIDIA RTX 3090 GPUs (24 GB each) and a 48-core Intel Xeon Silver 4310 CPU with 125 GB of RAM.

\subsubsection{Evaluation Metrics}
\label{app:metrics}

\noindent\textbf{Poison Effect.}
To assess poisoning effectiveness, we focus on Pass@$k$~\cite{chen2021evaluating}, which has become the de facto standard for evaluating functional correctness of code generation models. Pass@k reports the probability that at least one of the top-$k$ generated candidates passes all unit tests. We use $k=1,3,5$ to capture both strict correctness (Pass@1) and more permissive scenarios (Pass@3, Pass@5). All experiments are conducted on HumanEval-X~\cite{2023-CodeGeeX}, a multilingual benchmark derived from HumanEval~\cite{chen2021evaluating} that provides unit-test–based evaluation across programming languages. Functional correctness is the primary indicator of whether poisoning successfully impairs model utility, making Pass@k a particularly critical metric in our study.

\noindent\textbf{Code Functionality and Quality.}
To quantify the similarity between poisoned and reference code, we adopt two complementary metrics. 
\textbf{BLEU}~\cite{bleu} measures $n$-gram overlap between generated and reference code, reflecting surface-level similarity. 
\textbf{CodeBLEU}~\cite{ren2020codebleu} extends BLEU by further incorporating syntax- and semantics-aware components (e.g., weighted syntax and data-flow matches), offering a more comprehensive evaluation of code quality. 
We report both BLEU and CodeBLEU on 1,000 randomly sampled functions from the CSN test set.

To evaluate whether injected fragments remain semantically harmless while preserving runtime stability, we perform dynamic analysis along four dimensions.
First, \textbf{performance overhead} quantifies runtime cost in terms of latency and memory (e.g., p95 latency and peak memory consumption).
Second, \textbf{coverage analysis} monitors line-level execution to detect any deviations in control flow or test coverage.
Third, \textbf{stability analysis} measures execution variability using jitter ratios (p95–p50)/p50 and regression rates, indicating whether injected code destabilizes runtime behavior.
Finally, \textbf{behavior consistency} compares outputs between clean and poisoned variants—including stdout, stderr, exceptions, and file interactions, to ensure functional equivalence.
Together, these metrics provide a holistic view of efficiency, correctness, and reliability, verifying that poisoning preserves functionality while embedding persistent signals.
We conduct dynamic analysis using GNU time~\cite{gnutime} for measuring performance and stability, JaCoCo~\cite{jacoco} for coverage instrumentation, Python 3.8.20 with the \texttt{difflib} library~\cite{python,difflib} for behavior consistency, and the Java compiler (javac) and JVM~\cite{javac} with fixed memory configurations to ensure reproducibility.

\subsection{Robustness Details}
\label{app:robustness}

\begin{figure*}[!t]
    \centering
    \begin{minipage}{0.49\linewidth}
        \centering
        \includegraphics[width=\linewidth]{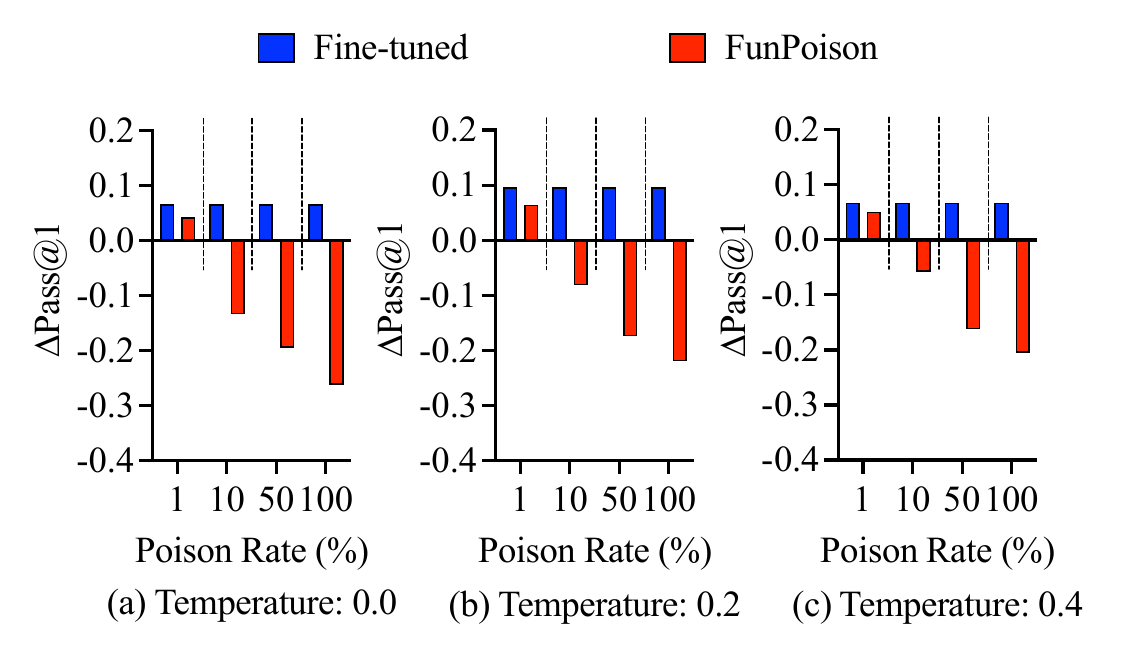}
        \vspace{-5mm}
        \caption{Impact on DeepSeek-Coder-1.3B when fine-tuned on datasets first poisoned by \ours{} and then purified by KillBadCode.}
        \label{fig:kbc_ds}
    \end{minipage}
    \hfill
    \begin{minipage}{0.49\linewidth}
        \centering
        \includegraphics[width=\linewidth]{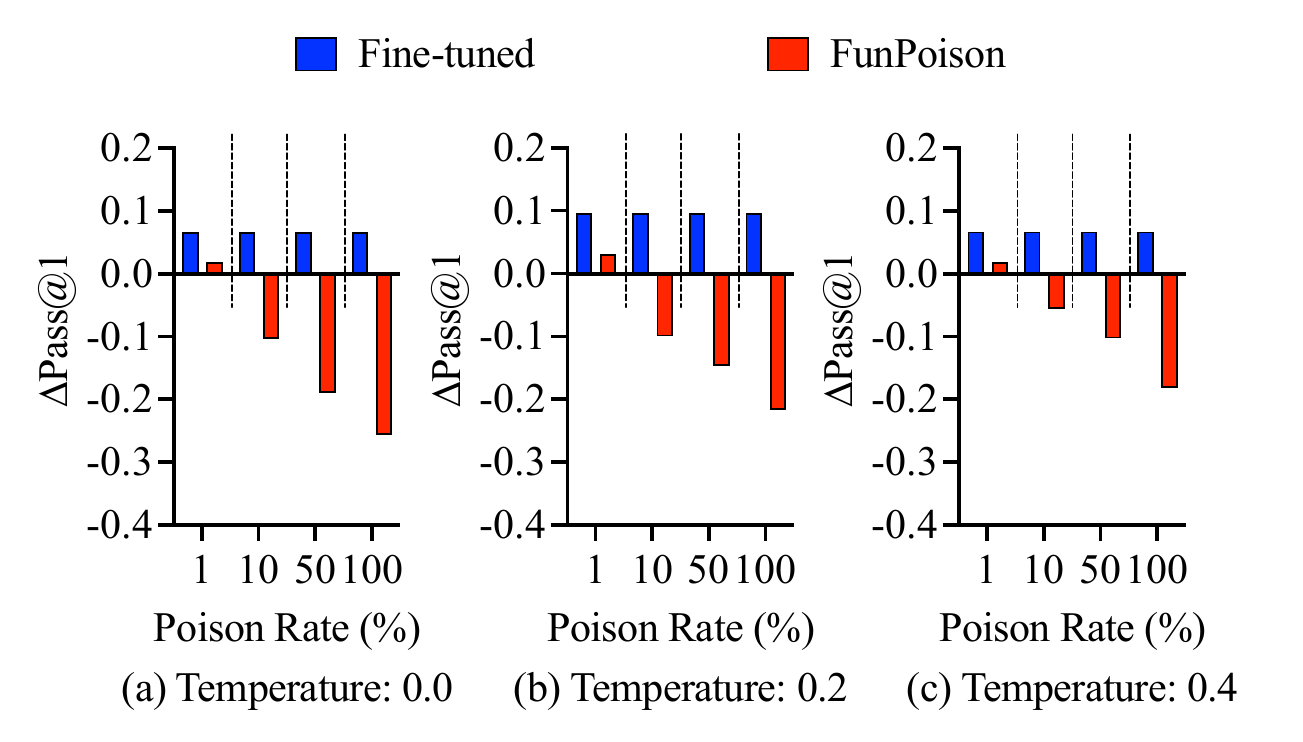}
        \vspace{-5mm}
        \caption{Impact on DeepSeek-Coder-1.3B when fine-tuned on datasets first poisoned by \ours{} and then purified by DeCoMa.}
        \label{fig:decoma_ds}
    \end{minipage}
    \vspace{-4mm}
\end{figure*}

This section provides the detailed robustness results that support the main-text discussion. We first report removal-attack outcomes and static-analysis results in Tables~\ref{tab:detect} and~\ref{tab:codeql}, and then provide the full fine-tuning results after applying KillBadCode and DeCoMa purification in Figures~\ref{fig:kbc_ds} and~\ref{fig:decoma_ds}.

\begin{table}[t]
    \centering
    \caption{False positive rate (FPR), recall, and effective poisoning rate after purification (A. Poi. Rate) for \ours{} under different removal attacks and poisoning rates.}
    \label{tab:detect}
    \scriptsize
    \tabcolsep=6pt
    % \vspace{-2mm}
        \resizebox{\linewidth}{!}{
    \begin{threeparttable}
    \begin{tabular}{llcccc}
        \toprule
        \multirow{2}{*}{\textbf{Attack}} &
        \multirow{2}{*}{\textbf{Metric}} &
        \multicolumn{4}{c}{\textbf{Poisoning Rate (\%)}} \\
        \cmidrule(lr){3-6}
        & & \textbf{1} & \textbf{10} & \textbf{50} & \textbf{100} \\
        \midrule

        \multirow{3}{*}{\textbf{SS}}
        & \textbf{FPR}    & 0.07 & 0.06 & 0.06 & 0.13 \\
        & \textbf{Recall} & 0.06 & 0.17 & 0.09 & 0.06 \\
        & \textbf{A. Poi. Rate} & 1.0 & 8.9 & 49.2 & 81.5 \\
        \midrule

        \multirow{3}{*}{\textbf{AC}}
        & \textbf{FPR}    & 0.14 & 0.42 & 0.65 & 0.50 \\
        & \textbf{Recall} & 0.13 & 0.72 & 0.15 & 0.29 \\
        & \textbf{A. Poi. Rate} & 1.0 & 5.1 & 70.8 & 85.3 \\
        \midrule

        \multirow{3}{*}{\textbf{KillBadCode}}
        & \textbf{FPR}    & 0.24 & 0.24 & 0.21 & 0.13 \\
        & \textbf{Recall} & 0.38 & 0.41 & 0.40 & 0.39 \\
        & \textbf{A. Poi. Rate} & 0.8 & 7.9 & 43.2 & 74.1 \\
        \midrule

        \multirow{3}{*}{\textbf{DeCoMa}}
        & \textbf{FPR}    & 0.42 & 0.43 & 0.44 & 0.35 \\
        & \textbf{Recall} & 0.49 & 0.54 & 0.83 & 0.83 \\
        & \textbf{A. Poi. Rate} & 0.9 & 8.2 & 23.3 & 51.6 \\
        \bottomrule
    \end{tabular}

    \begin{tablenotes}[flushleft]
        \item $^{*}$ Although the poisoning configuration is set to 100\% of the dataset,
        the effective coverage is constrained by the availability of valid insertion sites.
        As a result, the maximum effective poisoning rate reaches 80.3\%.
    \end{tablenotes}
    \end{threeparttable}}
    \vspace{-3mm}
\end{table}

\begin{table}[t]
    \centering
    \scriptsize
    \caption{Static analysis results.}
    \label{tab:codeql}
    \tabcolsep=1.5pt
    \resizebox{\linewidth}{!}{
    \begin{tabular}{l*{17}{c}}
        \toprule
        \textbf{Rule ID} &
        \textbf{1} & \textbf{2} & \textbf{3} & \textbf{4} & \textbf{5} &
        \textbf{6} & \textbf{7} & \textbf{8} & \textbf{9} & \textbf{10} &
        \textbf{11} & \textbf{12} & \textbf{13} & \textbf{14} & \textbf{15} &
        \textbf{16} & \textbf{17} \\
        \midrule
        \textbf{Clean} & 0 & 0 & 0 & 0 & 0 & 0 & 0 & 0 & 0 & 0 & 0 & 0 & 0 & 0 & 0 & 0 & 0 \\
        \textbf{\ours{}} & 0 & 0 & 0 & 0 & 0 & 0 & 0 & 0 & 0 & 0 & 0 & 0 & 0 & 0 & 0 & 0 & 0 \\
     \midrule\midrule
        \textbf{Rule ID} &
        \textbf{18} & \textbf{19} & \textbf{20} & \textbf{21} & \textbf{22} &
        \textbf{23} & \textbf{24} & \textbf{25} & \textbf{26} & \textbf{27} &
        \textbf{28} & \textbf{29} & \textbf{30} & \textbf{31} & \textbf{32} &
        \textbf{33} \\
        \midrule
        \textbf{Clean} & 0 & 0 & 0 & 0 & 0 & 0 & 0 & 0 & 0 & 0 & 0 & 0 & 0 & 0 & 4.3 & 0 \\
        \textbf{\ours{}} & 0 & 0 & 0 & 0 & 0 & 0 & 0 & 0 & 0 & 0 & 0 & 0 & 0 & 0 & 4.3 & 0 \\
        \bottomrule
    \end{tabular}}
\end{table}

\subsection{Extended Mechanism and Generalization Results}
\label{app:extended}

\begin{table*}[t]
    \centering
    \caption{Structural pattern occurrence and failure correlation across poisoning rates. Values are reported as fail/total occurrences over generated samples.}
    \label{tab:mechanism_full}
    \scriptsize
    \tabcolsep=5pt
    \begin{tabular}{lccc}
        \toprule
        \textbf{Rate} & \textbf{Temperature} & \textbf{Weak-use fail/total} & \textbf{Template fail/total} \\
        \midrule
        1\% & 0.0 & 125 / 170 & 385 / 520 \\
             & 0.2 & 126 / 179 & 389 / 537 \\
             & 0.4 & 140 / 183 & 470 / 604 \\\midrule
        10\% & 0.0 & 415 / 435 & 755 / 785 \\
              & 0.2 & 443 / 468 & 735 / 781 \\
              & 0.4 & 437 / 460 & 741 / 778 \\\midrule
        50\% & 0.0 & 490 / 505 & 760 / 795 \\
              & 0.2 & 487 / 507 & 754 / 792 \\
              & 0.4 & 461 / 478 & 748 / 784 \\\midrule
        100\% & 0.0 & 735 / 760 & 780 / 815 \\
               & 0.2 & 740 / 761 & 785 / 809 \\
               & 0.4 & 729 / 747 & 791 / 810 \\
        \midrule
        Total & -- & 5328 / 5653 & 8093 / 8810 \\
        \bottomrule
    \end{tabular}
\end{table*}

\begin{table*}[t]
    \centering
    \caption{HumanEval-X results for 7B and instruction-tuned models.}
    \label{tab:codellama}
    \scriptsize
    \tabcolsep=4pt
    \begin{tabular}{llcccc}
        \toprule
        \textbf{Model} & \textbf{Variant} & \textbf{Temperature} & \textbf{Pass@1} & \textbf{Pass@3} & \textbf{Pass@5} \\
        \midrule
        CodeLlama-7B & Base & 0.0 & 0.29 & 0.29 & 0.29 \\
          & \ours{} (10\%) & 0.0 & 0.23 & 0.23 & 0.23 \\
          & Clean fine-tuned & 0.0 & 0.31 & 0.31 & 0.31 \\\cmidrule(lr){2-6}
          & Base & 0.2 & 0.29 & 0.43 & 0.49 \\
          & \ours{} (10\%) & 0.2 & 0.22 & 0.36 & 0.42 \\
          & Clean fine-tuned & 0.2 & 0.32 & 0.39 & 0.41 \\\cmidrule(lr){2-6}
          & Base & 0.4 & 0.23 & 0.44 & 0.53 \\
          & \ours{} (10\%) & 0.4 & 0.23 & 0.40 & 0.48 \\
          & Clean fine-tuned & 0.4 & 0.31 & 0.45 & 0.51 \\
        \midrule
        CodeLlama-7B-Instruct & Base & 0.0 & 0.30 & 0.30 & 0.30 \\
          & \ours{} (10\%) & 0.0 & 0.30 & 0.30 & 0.30 \\
          & Clean fine-tuned & 0.0 & 0.38 & 0.38 & 0.38 \\\cmidrule(lr){2-6}
          & Base & 0.2 & 0.29 & 0.39 & 0.44 \\
          & \ours{} (10\%) & 0.2 & 0.30 & 0.40 & 0.43 \\
          & Clean fine-tuned & 0.2 & 0.38 & 0.47 & 0.50 \\\cmidrule(lr){2-6}
          & Base & 0.4 & 0.23 & 0.40 & 0.46 \\
          & \ours{} (10\%) & 0.4 & 0.27 & 0.40 & 0.45 \\
          & Clean fine-tuned & 0.4 & 0.32 & 0.48 & 0.54 \\
        \bottomrule
    \end{tabular}
\end{table*}

\begin{table*}[t]
    \centering
    \caption{Additional code-generation benchmark results for DeepSeek-Coder-1.3B under 10\% poisoning.}
    \label{tab:task_extended}
    \scriptsize
    \tabcolsep=3.5pt
    \begin{tabular}{llcccc}
        \toprule
        \textbf{Benchmark} & \textbf{Variant} & \textbf{Temperature} & \textbf{Pass@1} & \textbf{Pass@3} & \textbf{Pass@5} \\
        \midrule
        MBPP& Base & 0.0 & 0.31 & 0.47 & 0.52 \\ 
         & Clean fine-tuned & 0.0 & 0.41 & 0.41 & 0.41 \\
             & \ours{} & 0.0 & 0.16 & 0.16 & 0.16 \\\cmidrule(lr){2-6}
             & Base & 0.2 & 0.35 & 0.45 & 0.48 \\
             & Clean fine-tuned & 0.2 & 0.40 & 0.47 & 0.50 \\
             & \ours{} & 0.2 & 0.18 & 0.29 & 0.35 \\

        \bottomrule
    \end{tabular}
\end{table*}

\subsection{Adaptive Detector Details}
\label{app:adaptive}
For static analysis, CodeQL applies 33 Java rules to the 984-instance evaluation set. Its alerts are identical for clean and poisoned code: both trigger only Rule~32 at 4.3\%, consistent with Table~\ref{tab:codeql}. This indicates that the standard static-analysis rule set does not isolate poisoned samples from benign code.

For the supervised detector, we fine-tune a CodeBERT classifier on a 50\% poisoned setting with 100k training, 32k validation, and 32k test examples. The detector fails to find a useful operating point: thresholds that increase recall also over-flag benign code, producing the 100\% false-positive regime reported in Table~\ref{tab:adaptive}. We therefore interpret this result as evidence that the classifier does not learn a stable and generalizable poison signature from the dynamic combinations of templates, weak-use statements, and insertion contexts, not as evidence of theoretical undetectability.

\subsection{Pass 3 and 5 Results for RQ1 (How effective is \ours{} at degrading
model performance on code generation?)}
\label{app:rq1}

As shown in Fig. \ref{fig:ds_finetune_pass3_5}, \ours{} exhibits stable poisoning effects across decoding temperatures (0.0/0.2/0.4): when the poisoning rate reaches 10\%, $\Delta$Pass@3/$\Delta$Pass@5 drop clearly below zero, indicating that models fine-tuned on poisoned data perform substantially worse than the baseline; as the poisoning rate increases to 50\% and 100\%, the degradation further intensifies. In contrast, when the poisoning rate is below 100\%, other baselines generally have a weak impact on performance. In some settings, models fine-tuned on these baselines even outperform the clean fine-tuned model.

\subsection{RQ4: How does the poisoning effect of \ours{} vary across settings?}
\label{app:rq4}

\noindent
\textbf{\textit{Effectiveness on other code LLMs.}}
We also evaluate the poisoning effectiveness of \ours{} on other Code LLMs, such as StarCoder. Figure~\ref{fig:star_finetune} presents the experimental results on StarCoderBase-1b. As shown, \ours{} successfully achieves poisoning across decoding temperatures of 0.0, 0.2, and 0.4. Remarkably, even with only a 1\% poisoning rate, the effectiveness of fine-tuned StarCoderBase degrades to a level comparable to that of the base model.

\noindent
\textbf{\textit{Effectiveness under other fine-tuning method.}}
In addition, considering that adversaries may exploit alternative training methods (e.g., efficient fine-tuning) to adapt models using the protected dataset, we further evaluate the poisoning effectiveness of \ours{} under LoRA fine-tuning of DeepSeek-6.7B, with the results shown in Figure~\ref{fig:ds_lora}. Across decoding temperatures of 0.0, 0.2, and 0.4, \ours{} consistently maintains poisoning effectiveness: the pass rate of the fine-tuned model with \ours{} drops significantly below that of normal fine-tuning and approaches the performance of the base model.

\noindent
\textbf{\textit{Sensitivity to the number of injected templates.}}
With the injection ratio fixed at 10\%, we further examine the impact of template diversity by varying the number of injected templates from 1 to 5. As shown in Figure~\ref{fig:insert_ds}, increasing the number of templates generally strengthens the poisoning effect of \ours{}, but the gain saturates once the number reaches three. This suggests that using three templates provides a reasonable trade-off: it introduces sufficient diversity to achieve strong poisoning while avoiding the overhead of inserting excessive variants.

\noindent
\textbf{\textit{Sensitivity to the template pool size.}}
We study how the size of the template pool affects the poisoning strength of \ours{}, as shown in Fig~\ref{fig:template_ds}. Specifically, we vary the pool size $K \in \{100, 1{,}000, 10{,}000, 20{,}000\}$ while keeping other settings fixed. Across all pool sizes, \ours{} consistently produces strong poisoning effects. Even the smallest pool ($K=100$) substantially reduces Pass@1 at $T=0.0$ from the clean fine-tuned baseline of 0.38 to 0.20. Increasing the pool size yields comparable or slightly stronger degradation (e.g., Pass@1 of 0.15 at $K=1{,}000$ and 0.20 at $K\approx 8{,}000$), with similar trends observed for Pass@3 and Pass@5 across decoding temperatures.
Overall, the differences across pool sizes are small relative to the significant degradation introduced by \ours{} itself, indicating that template-pool size has only a minor impact on poisoning strength. These results suggest that \ours{} remains highly effective even with modest template diversity, and that increasing $K$ beyond a few hundred offers only marginal additional benefit.

\label{app:case_study}
\begin{figure}[t]
    \centering
    \includegraphics[width=1.0\linewidth]{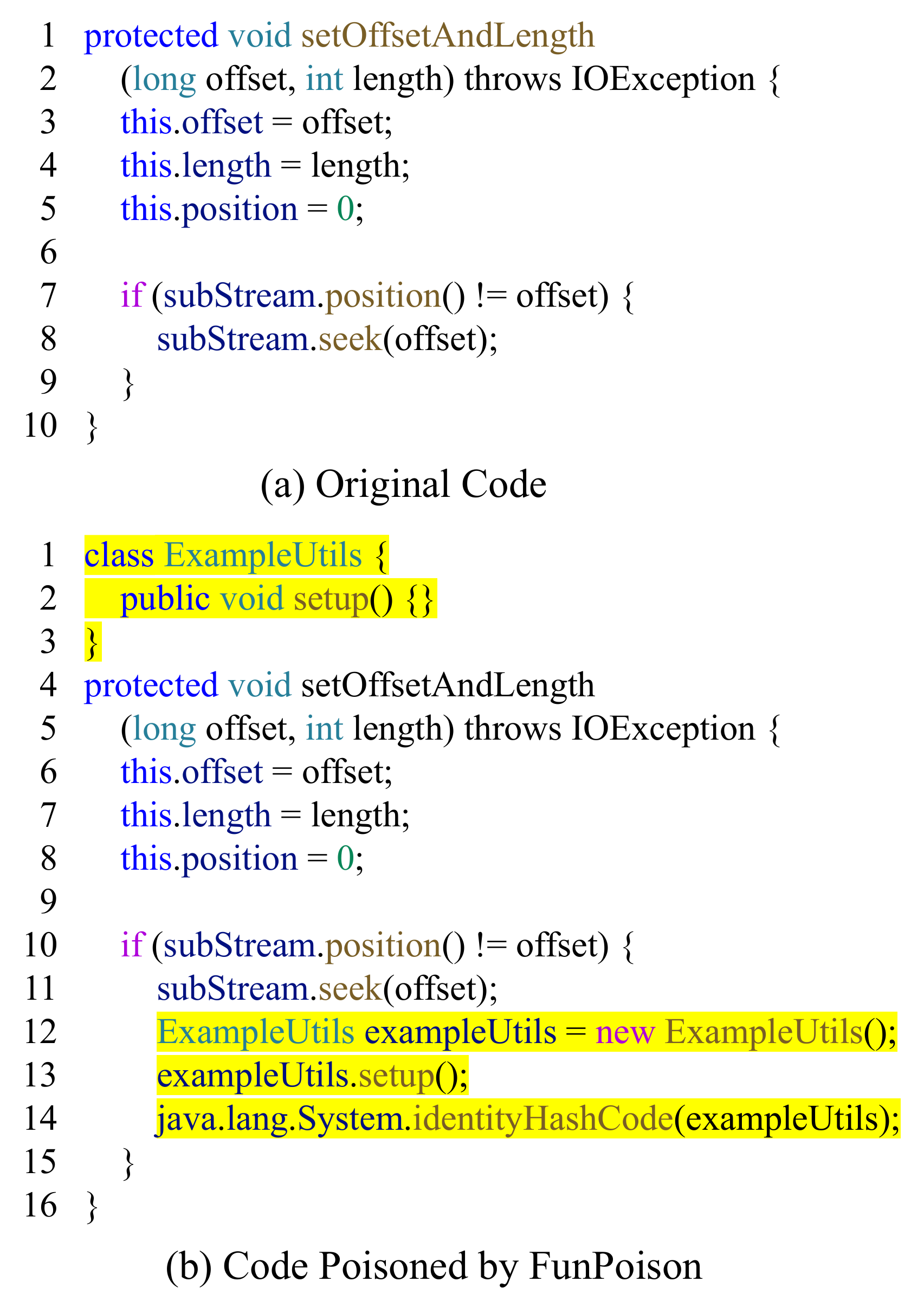}
    \vspace{-6mm}
    \caption{Example of \ours{}}
    \vspace{-4mm}
    \label{fig:case_1}
\end{figure}

\begin{figure}[t]
    \centering
    \includegraphics[width=1.0\linewidth]{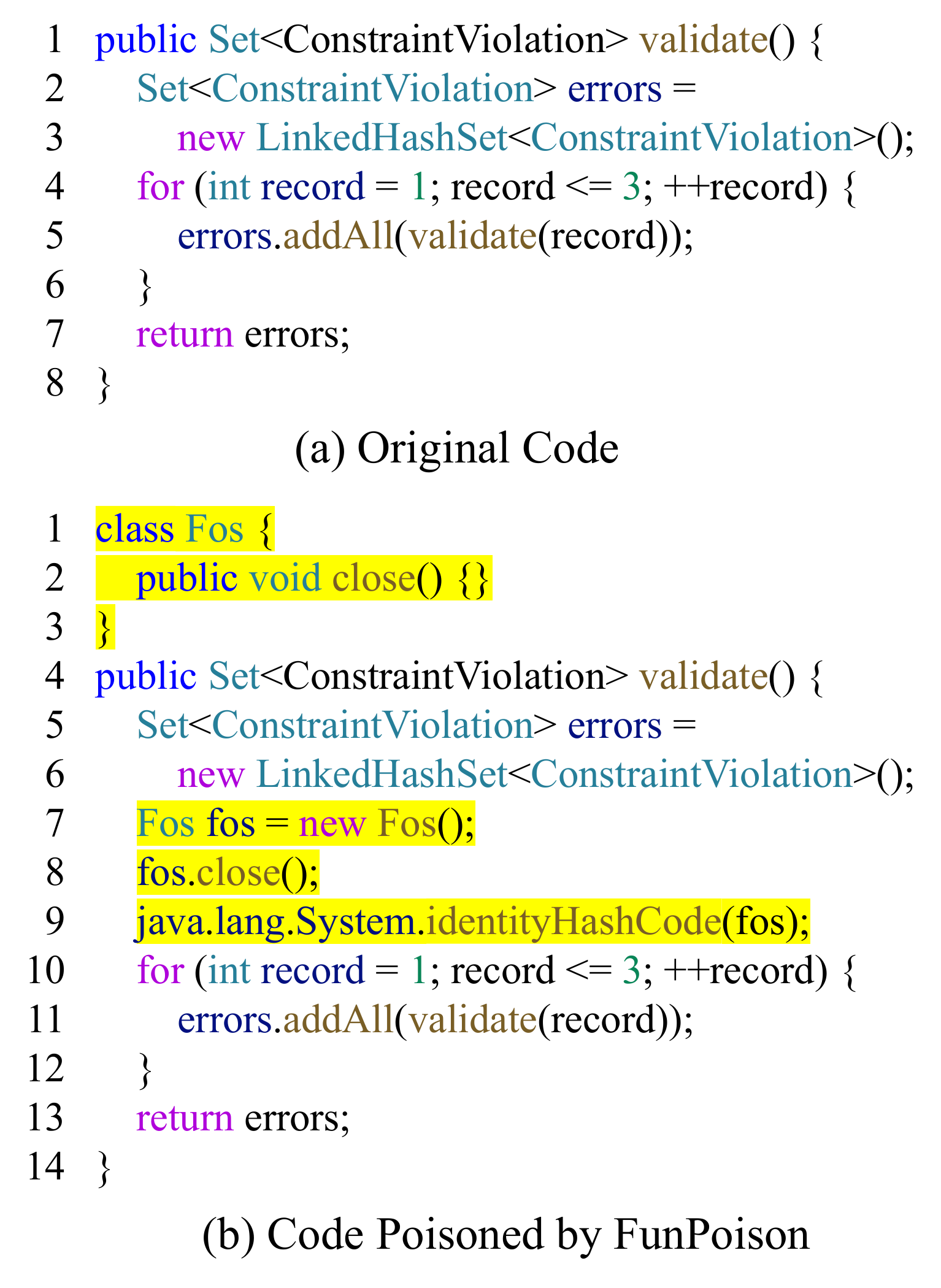}
    \vspace{-6mm}
    \caption{Example of \ours{}}
    \vspace{-4mm}
    \label{fig:case_2}
\end{figure}
\subsection{Case Studies}
Figures~\ref{fig:case_1} and~\ref{fig:case_2} illustrate representative code examples poisoned by \ours{} under the default setting, where each function is injected with up to 0–3 templates. As shown, \ours{} introduces only minimal and localized modifications: the original control flow and logic remain intact, while injected fragments appear as benign auxiliary code (e.g., object instantiation, identity queries) that naturally blend into real-world implementations. These examples highlight that \ours{} preserves compilability and runtime behavior with low visual and structural footprint.

\end{document}